\documentclass[]{jfm}

\usepackage{graphicx}
\usepackage{graphbox}
\usepackage{newtxtext}
\usepackage{newtxmath}
\usepackage{mathtools}
\usepackage{natbib}
\usepackage{hyperref}
\hypersetup{
    colorlinks = true,
    urlcolor   = blue,
    citecolor  = black,
}

\usepackage{multirow}
\usepackage{accents}
\usepackage{comment}
\usepackage{subcaption}
\usepackage{xcolor} 
\usepackage{epstopdf, epsfig}
\usepackage{tikz}

\newcommand{\RomanNumeralCaps}[1]
\nolinenumbers
\graphicspath{{figures/}}
\usepackage{cleveref} 

\newcommand{\hg}{{\bf \hat{g}}}

\title{Long-wave instabilities of sloping stratified exchange flows}

\author{Lu Zhu\aff{1} 
	\corresp{\email{lz447@cam.ac.uk}}, Amir Atoufi\aff{1}, Adrien Lefauve\aff{1}, Rich R. Kerswell\aff{1}, P. F. Linden\aff{1}}

\affiliation{\aff{1}Department of Applied Mathematics and Theoretical Physics, University of Cambridge, Centre for Mathematical Sciences, Wilberforce Road, Cambridge CB3 0WA, UK
}

\begin{document}

\maketitle

\begin{abstract}
We investigate the linear instability of two-layer stratified shear flows in a sloping two-dimensional channel, subject to non-zero longitudinal gravitational forces.
We reveal three previously unknown instabilities, distinct from the well-known Kelvin-Helmholtz Instability (KHI) and Holmboe Wave Instability (HWI), in that they have longer wavelengths (of the order of 10 to $10^3$ shear-layer depths) and often slower growth rates. Importantly, they can grow in background flows with gradient Richardson number $\gg 1$, which offers a new mechanism to sustain turbulence and mixing in strongly stratified flows.
These instabilities are shown to be generic and relatively insensitive to Reynolds number $\Rey$, Prandtl number $\Pran$, base flow profile, and boundary conditions. 
The nonlinear evolution of these instabilities is investigated through a forced direct numerical simulation, in which the background momentum and density are sustained. The growth of long unstable waves in background flows initially stable to short wave instabilities causes a decrease in the local gradient Richardson number. This leads to local nonlinear processes that result in small-scale overturns resembling Kelvin-Helmholtz billows. Our results establish a new energy exchange pathway, where the mean kinetic energy of a strongly stratified flow is extracted  
by primary unstable long waves and secondary short waves, and subsequently dissipated into internal energy.

\end{abstract}


\section{Introduction}

The study of stratified flows has attracted considerable attention over the past few decades due to their importance in many environmental and industrial processes. In the oceans, stratification occurs due to differences in salinity and/or temperature, leading to mostly stably stratified flows. Turbulence in these flows plays a significant role in the transport of momentum and mass and is crucial in shaping the global climate~\citep{linden1979mixing,riley2000fluid,gregg_mixing_2018,colm2020open}. 
An interesting open question concerns the maintenance of turbulence and its associated irreversible turbulent mixing under strong stable stratification, which tends to suppress turbulence.

When stratification is relatively weak, stably-stratified flows can be linearly unstable. It is well known that linear shear instabilities, such as Kelvin-Helmholtz instability (KHI)~\citep{hazel1972numerical, Smyth1988} and Holmboe wave instability (HWI)~\citep{holmboe1962behavior}, can cause transition of a laminar stratified flow to turbulence, inducing strong mass and momentum transport~\citep{caulfield2021layering}. Over the past 50 years, numerous studies have been carried out to understand these instabilities and their relation to mixing~\citep{thorpe1968method, Smyth1988, carpenter2010holmboe, salehipour_peltier_mashayek_2015, zhou2017diapycnal}. In most of these studies, the density isopycnals are perpendicular to the direction of gravity, which does not explicitly drive the flows.

However, in many natural systems, density isopycnals are not exactly perpendicular to gravity, in which case nonzero streamwise gravity forces come into play and may partially drive the flow. One notable example is the internal tide interacting with the sloping bottom topography of the oceanic continental shelf~\citep{garrett2007internal}. At a critical slope, the internal tide provides an additional energy production pathway that leads to turbulent mixing of temperature, salinity, and other tracers~\citep{gayen2010turbulence}.
Similarly, many engineering flows occur along an inclined boundary. Examples can be found in building ventilation systems~\citep{linden1999fluid}, where indoor/outdoor air is often exchanged through inclined ventilation ducts, producing mixing and dispersion of heat and indoor pollutants.
In gas-cooled nuclear reactors, carbon dioxide and air are exchanged through inclined coolant ducts, which can result in the depressurization and damage of the reactors in case of failure~\citep{leach1975investigation,mercer1975experimental}.

Studies on the influence of longitudinal gravitational forcing on the onset of turbulence in stratified exchange flows remain limited. One notable recent body of work is the Stratified Inclined Duct (SID) experiment~\citep{meyer2014stratified,lefauve2019regime,lefauve2020buoyancy}. These studies investigated the transition and turbulent mixing of the exchange flow in an inclined duct that connected two reservoirs with fluids at different densities or temperatures. To understand the mechanism of transition in SID, \citet{lefauve2018structure} conducted a linear stability analysis using a base state extracted from the SID experiment.
Subsequently, \citet{ducimetiere2021effects} systematically investigated the three-dimensional unstable modes in inclined ducts, focusing on the effects of side wall confinement. 
These studies focused primarily on HWI (and secondarily on KHI), which have wavelengths comparable to the thickness of the shear layers. Interestingly, \citet{ducimetiere2021effects} observed a secondary instability at significantly longer wavelengths than KHI and HWI and attributed it to the effect of the inclination angle. Recently, \citet{ZhuAtoufi2022_II} studied the mechanism of transition by applying shallow water equations as a diagnostic tool to analyse a new numerical database of SID~\citep{ZhuAtoufi2022}. They suggested that the instability of long shallow water waves (long-wave KHI in the presence of top and bottom solid boundaries) may cause turbulence in the SID. Although the longitudinal gravitational forcing was included in the numerical simulation data, it was not included explicitly in the shallow water model. 

In this paper, we explore explicitly the impact of longitudinal gravitational forces on the instability of long waves and on potential new pathways toward turbulence, restricting ourselves to a two-dimensional geometry. In  \S~\ref{sec:LSA}, we examine the linear instabilities in inclined channels and conduct a thorough exploration of the parameter space. 
We identify three new distinct families of long-wave instabilities distinct from the well-known HWI and KHI, and map in parameter space these long-wave instabilities that dominate the flow. In  \S~\ref{sec:dns}, we then investigate the evolution of these new instabilities by conducting two-dimensional forced direct numerical simulations (DNS), and discuss their impact on turbulence and energy transfers. Finally, we conclude in \S~\ref{sec:conclusions}.

\section{Methodology}\label{sec:method}

	\subsection{Problem formulation and governing equations} \label{sec:gov-eqs}
%
%
\begin{figure}
	\centering		
	\includegraphics[align=c,width=.55\linewidth, trim=0mm 0mm 0mm 0mm, clip]{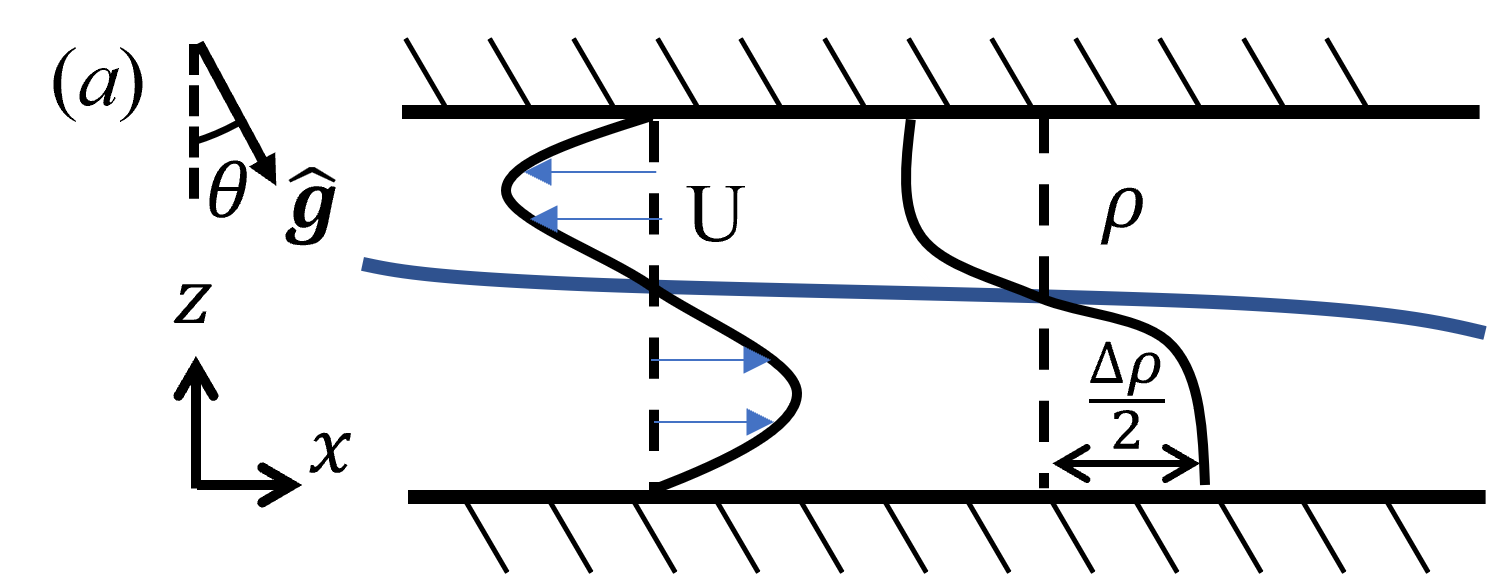}
	\includegraphics[align=c,width=.25\linewidth, trim=0mm 0mm 0mm 0mm, clip]{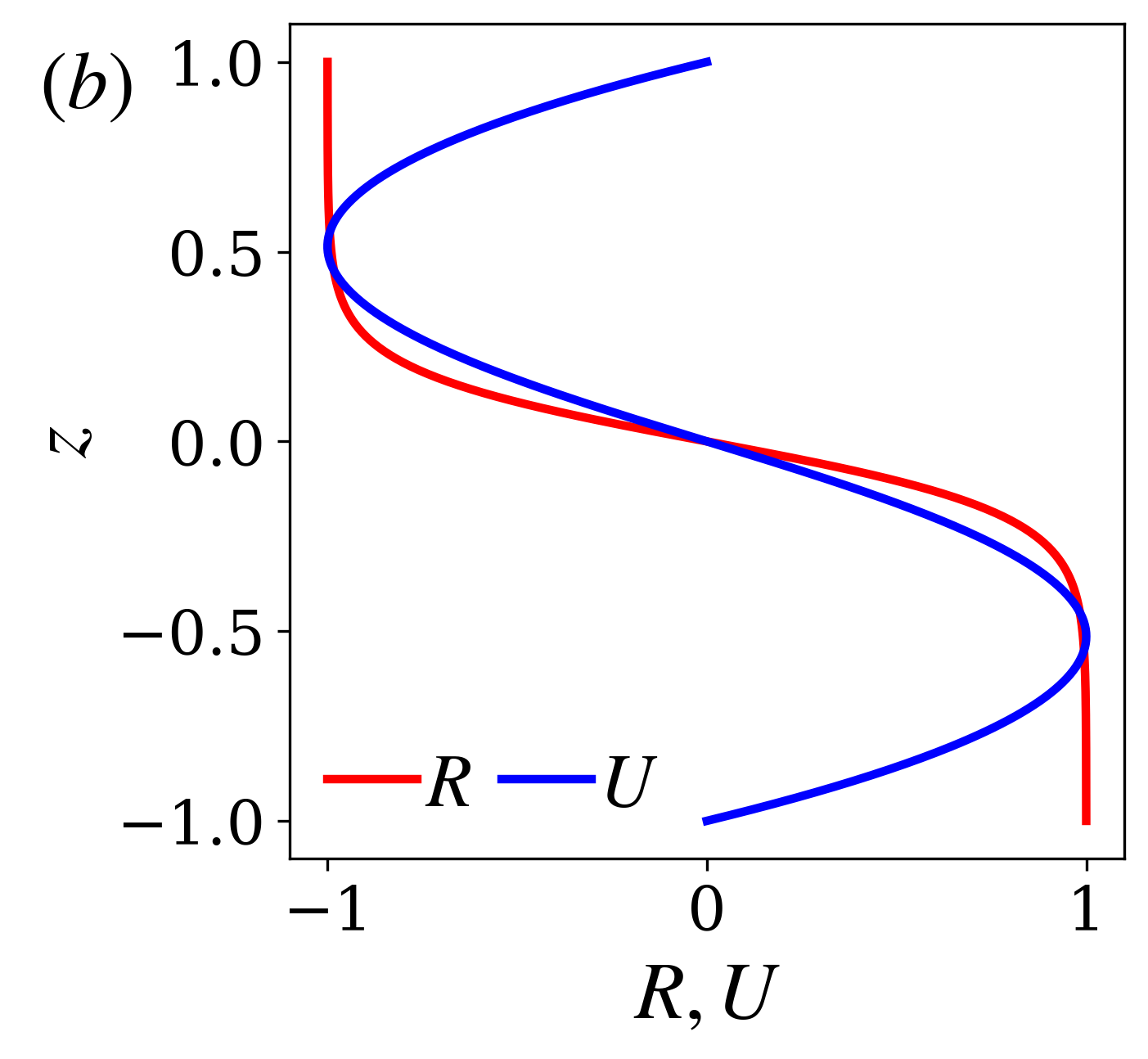}
	\caption{(a) Schematic of the two-dimensional shear flow in a stratified channel inclined at an angle $\theta$, and (b) base velocity $U(z)$ and density $R(z)$ profiles computed from (\ref{eq:LSA_R}) and (\ref{eq:Mot}). }
	\label{fig:geom}
\end{figure}

    In this section, we present the equations required for linear stability analysis (LSA) of a stratified exchange flow between two fluid layers having density $\rho_0 \pm\Delta \rho/2$ (where $\rho_0$ is the reference density and $0<\Delta \rho\ll \rho_0$ is the density difference) in a two-dimensional stratified inclined channel (SIC, see \cref{fig:geom}(a)).
    Following the SID experimental literature, lengths are nondimensionalized by the half-channel height $H^*$, velocity by the 
    buoyancy-velocity scale $U^* \equiv \sqrt{g^\prime H^*}$ (where $g^\prime=g\Delta \rho/\rho_0$ is the reduced gravity), 
    time by the advective time unit $H^*/U^*$, pressure by $\rho_0 U^{*2}$, and density variations around $\rho_0$ by $\Delta\rho/2$, respectively.
    The non-dimensional continuity, Navier-Stokes  and scalar governing equations under the Boussinesq approximation are
	\begin{eqnarray}
		\label{eq:ns_mass}%
		\boldsymbol{\nabla} \cdot \mathbf{u} &=& 0,%
		\\
		\label{eq:ns_mom}%
		\frac{\partial \mathbf{u}}{\partial t}+\mathbf{u}\cdot \boldsymbol{\nabla}\mathbf{u}  &=&
		- \boldsymbol{\nabla}p + \frac{1}{\mathrm{Re}} \nabla^{2}\mathbf{u} 
		+ \mathrm{Ri} \, \rho \, \hg,%
		\\
		\label{eq:ns_den}%
		\frac{\partial \rho}{\partial t}+\mathbf{u}\cdot \boldsymbol{\nabla}\rho &=&
		\frac{1}{\mathrm{\Rey \ \Pran}} \nabla^{2}\rho,%
	\end{eqnarray}
	where
	$\mathbf{u}=(u,v,w)$ is the non-dimensional velocity in the three-dimensional coordinate system $\mathbf{x}=(x,y,z)$, where $x$- ,$y$- ,$z$-axis are the longitude, spanwise and wall-normal direction of the channel respectively. In this coordinate system gravity $\mathbf{g}$ is pointing downward at a angle $\theta$ to the $-z$ axis, i.e., $\mathbf{g}=g \, 
  \hg=g\,[\sin{\theta}, 0, -\cos{\theta}]$, and $p$ and $\rho$ are the non-dimensional pressure and density, respectively. The dimensionless parameters are the Reynolds number Re $\equiv H^*U^*/\nu$ ($\nu$ is the kinematic viscosity), the Prandtl number $\Pran\equiv \nu/\kappa$ ($\kappa$ is the scalar diffusivity), and Richardson number Ri $ \equiv  g^\prime H^\ast/(2U^{\ast})^2=1/4$ (fixed here because of the buoyancy velocity scale).

	\subsection{Formulation of linear stability analysis}

    We now apply a linear stability analysis (LSA) ~\citep{drazin_reid_2004,smyth2019instability} to the SIC, noting that in agreement with Squire's theorem,~\citet{lefauve2018structure,ducimetiere2021effects} have shown that the fastest-growing mode is two dimensional (2D).  We impose infinitesimal 2D perturbations to a 1D base state. 
    The velocity, density, and pressure fields are thus decomposed as
    \begin{eqnarray}
        \label{eq:u_decomp}
        \boldsymbol{u}&=&\boldsymbol{U}+\boldsymbol{u}^\prime=[U(z),0,0]+[u^\prime,0,w^\prime],
        \\
        \label{eq:p_decomp}
        p &=& P(z)+p^\prime,
        \\
        \label{eq:rho_decomp}
        \rho &=& R(z) +\rho^\prime,
    \end{eqnarray}
    where capital letters and superscript prime represent the mean and perturbation components of quantities, respectively. A normal mode perturbation of the form
    \begin{equation}\label{eq:perturbation-form}
        \phi(x,z,t) = \hat{\phi}(z)\exp{(ikx+\eta t)},
    \end{equation}
    is adopted. The base flows are obtained by solving for the numerical solution of the laminar exchange flow following~\citet{thorpe1968method}, which will be introduced in \S\ref{sec:base}.
    Substituting (\ref{eq:u_decomp})-(\ref{eq:rho_decomp}) into (\ref{eq:ns_mass})-(\ref{eq:ns_den}) and linearising yields the same system as \citet{lefauve2018structure}, i.e.
   \begin{eqnarray} \label{eq:LS}
	\eta \left[\begin{array}{cc}
		\Delta & \mathsfbi{0} \\
		\mathsfbi{0} & \mathsfbi{I} 
	\end{array}\right]  \left[\begin{array}{c}
		\widehat{w} \\
		\widehat{\rho}
	\end{array}\right]=\left[\begin{array}{cc}
		\mathcal{L}_{ww} & \mathcal{L}_{w \rho} \\
		\mathcal{L}_{\rho w} &  \mathcal{L}_{\rho \rho}
	\end{array}\right] \left[\begin{array}{c}
		\widehat{w} \\
		\widehat{\rho}
	\end{array}\right] , 
	\end{eqnarray}
	where $\mathsfbi{0}$ and $\mathsfbi{I}$ are the zero and identity matrices, respectively and
	\begin{eqnarray}
  && \mathcal{L}_{ww}=-\mathrm{i}k U \Delta + \mathrm{i}k \mathscr{D}^2 U + \Rey^{-1} \Delta^2, \\ && \nonumber 
     \mathcal{L}_{w \rho}=Ri \left(k^2 \cos{\theta} -\mathrm{i} k \sin{\theta} \ \mathscr{D} \right), \\&& \nonumber
   \mathcal{L}_{\rho w} = -\mathscr{D}R, \\&& \nonumber
   \mathcal{L}_{\rho \rho} = -\mathrm{i} k U + \left(\Rey \ \Pran \right)^{-1} \Delta,
\end{eqnarray}
    where $\Delta=\mathscr{D}^2-k^2$ (the operator $\mathscr{D}=\partial/\partial z$ and $\mathscr{D}^2=\partial^2/\partial z^2$). At the top and bottom boundaries ($z=\pm 1$), no-slip and no-flux boundary conditions are applied for velocity and density, respectively. 
    We also demonstrate the negligible effect of choosing a free-slip boundary condition for velocity in \cref{sec:LSA_freeslip}. To obtain the unstable modes, we solve the linear system (\ref{eq:LS}) numerically using a second-order finite-difference discretization method described in \citet{smyth2019instability}. The spatial resolution is chosen based on the sharpness of the interface and is $(150, 150, 250, 400)$ grid points for $\Pran=(1$, $7$, $28$, $70)$, respectively. A sensitivity analysis for resolution ensured convergence of the results.

\subsection{Base flows}\label{sec:base}

The base state for density in our exchange flow is taken as a hyperbolic tangent (figure~\ref{fig:geom}(b))
\begin{equation} 
    R(z)=-\tanh(z/\delta)=-\tanh(2\sqrt{\Pran} \,z).
    \label{eq:LSA_R}
\end{equation}
The interfacial thickness is $\delta = 1/(2\sqrt{\Pran})$ to approximate the effect of diffusion~\citep{smyth_peltier_1991}. 
The typical model~\citep[e.g.]{Smyth1988} considers a shear layer driven by an arbitrary, controllable background shear. 
A similar procedure is applied to our SIC by modifying the laminar solution developed by \citet{thorpe1968method} and imposing a background body force $ \mathcal{F}=-\gamma Ri R$ (where $\gamma$ is a variable to control the magnitude of the force). This decouples the base velocity from the inclination angle in SIC, allowing for the exploration of the $U-\theta$ space, as if being influenced by arbitrary external tidal forces or pressure gradients.  
The mean velocity profile $U(z)$ of the steady laminar exchange flow is obtained by integrating the 2D momentum equation
\begin{eqnarray}\label{eq:Mot}
	-\frac{\p P}{\p x}+Ri \sin{\theta} R + \frac{1}{\Rey} \frac{\p ^2 U}{\p z^2}+\mathcal{F}=0,
\end{eqnarray} 
where $-\p P/\p x= 0$ to satisfy the zero-flux condition of SIC.
This yields the following laminar base state for the forced SIC
\begin{equation} 
    \label{eq:Us_full}
	U(z) = -\ \Rey \ Ri (\sin{\theta}-\gamma)I(z)+c_1 z+c_2,
\end{equation}
where
\begin{eqnarray}
   I(z;\Pran)&=&\frac{z^2}{2}+\ln{2} \delta z + \dfrac{\delta^2\operatorname{Li}_2\left(-\mathrm{e}^{2 z/\delta}\right)}{2},
\end{eqnarray}
where $Li_2$ is the polylogarithm function of order 2. The constants $c_1$ and $c_2$ are computed given the no-slip boundary condition at the walls $U(z=\pm 1)=0$ and are
\begin{eqnarray}
    c_1 &=& \frac{1}{2} \Rey \ Ri \ (\sin{\theta}-\gamma) \left[I(1)-I(-1)\right],
	\\
	c_2 &=&  \frac{1}{2}  \Rey \ Ri \ (\sin{\theta}-\gamma) \left[I(1)+I(-1)\right]. 
\end{eqnarray} 
This solution $U(z)$ is sinusoidal-like (figure~\ref{fig:geom}(b)), much like those observed in experiments and simulations~\citep{lefauve2018structure,ZhuAtoufi2022}. The magnitude of the base velocity  depends on $\Rey$, $\theta$, and $\gamma$, while the shape depends more on $\delta$. 
In addition to the base state described by (\ref{eq:Us_full}), we also conducted a LSA with a $\tanh$-shape velocity profile in \cref{sec:LSA_freeslip}, to compare with the standard stratified free-shear layer model~\citep{Smyth1988}. These results were qualitatively consistent with those in the remainder of the paper, in terms of the existence of the same long- and short-wave families in SIC.

\section{Results: new families of linear instabilities in SIC}\label{sec:LSA}

Here we present the results from the LSA  of SIC.
We explore the parameter space of $\theta-\gamma$ and map out three new families of long-wave instabilities in addition to the well-known short-wavel HWI and KHI. We also investigate the impacts of $\Rey$ and $\Pran$ in order to further understand the importance of these newly discovered long waves in the laminar-turbulence transition.

\subsection{Five families of instabilities}\label{sec:param_space}

We first fix $(\Rey,\Pran,\mathrm{Ri})=(1000,7,0.25)$ and vary the inclination angle $\theta$ from $-10^{\circ}$ to $10^{\circ}$. When $\theta>0$, the SIC slopes downward, the streamwise gravity energises the mean flow and vice versa. We vary the forcing factor $\gamma$, on which two important physical quantities depend: the interfacial background  Richardson number $Ri_b$, defined as the gradient Richardson number of the background flow at the density interface $z=0$, i.e.,
\begin{equation}\label{eq:Ri_b}
    Ri_b\equiv Ri\frac{\partial R/\partial z}{(\partial U/\partial z)^2}\Big|_{z=0},
\end{equation}
and the mass flux (or flow rate of buoyancy), which is given by
\begin{equation}
    \label{eq:Qm}
    Q_m \equiv \frac{1}{2}\int_{-1}^1 \, R U \, dz.
\end{equation}
The Richardson number $Ri_b$ is an important measure of the relative importance of stratification compared with shear, which is critical for stratified shear flow stability~\citep{colm2020open}. The mass flux $Q_m$ is closely associated with the hydraulic control of exchange flows; a threshold value of $Q_m\approx 0.5$ indicates the emergence of an internal hydraulic jump  ~\citep{meyer2014stratified,lefauve2019regime} which \cite{ZhuAtoufi2022_II} demonstrated to be equivalent to a relatively long KHI (requiring the existence of a top and bottom boundaries).

\begin{figure}
	\centering		
	\includegraphics[width=.9\linewidth, trim=0mm 0mm 0mm 0mm, clip]{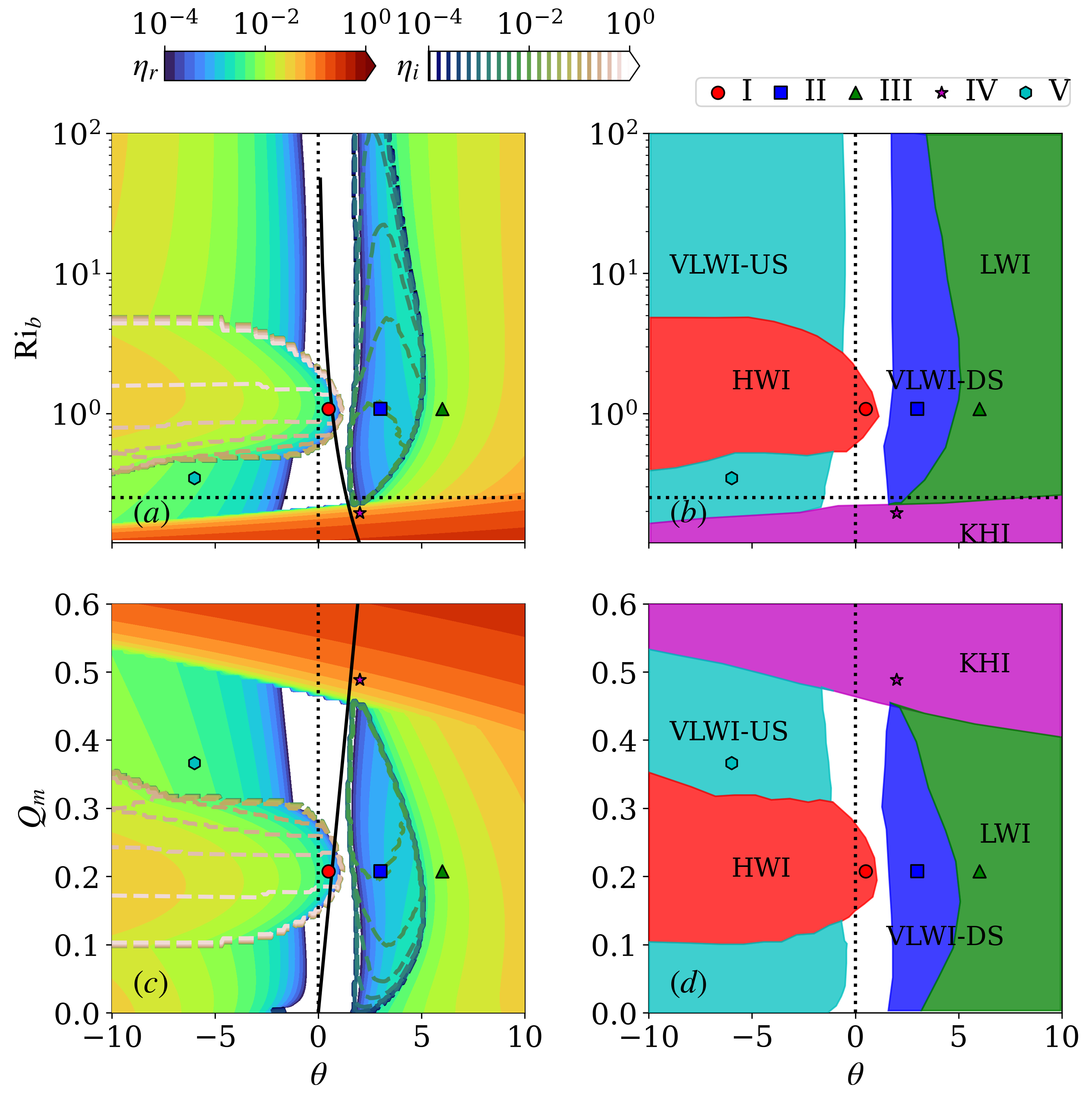}
	\caption{Parameter space projections of the fastest growing mode: (a) the growth rate $\eta_r$ (colours) and wave frequency $\eta_i$ (lines) and (b) the schematics of the $Ri_b-\theta$ parameter space; (c) the growth rate $\eta_r$ and wave frequency $\eta_i$ and (d) the schematics of $Q_m-\theta$ parameter space. Markers represent the five cases I, ..., V in Table \ref{tab:resol_DNS} for which the fastest growing mode is calculated. Black solid lines are the natural convective Thorpe base state (i.e. $\mathcal{F}=0$), and the horizontal dotted lines in (a) and (b) correspond to $Ri_b = 0.25.$ }
	\label{fig:Qm_ric_theta}
\end{figure}

\begin{figure}
	\centering		
	\includegraphics[width=.95\linewidth, trim=0mm 0mm 0mm 0mm, clip]{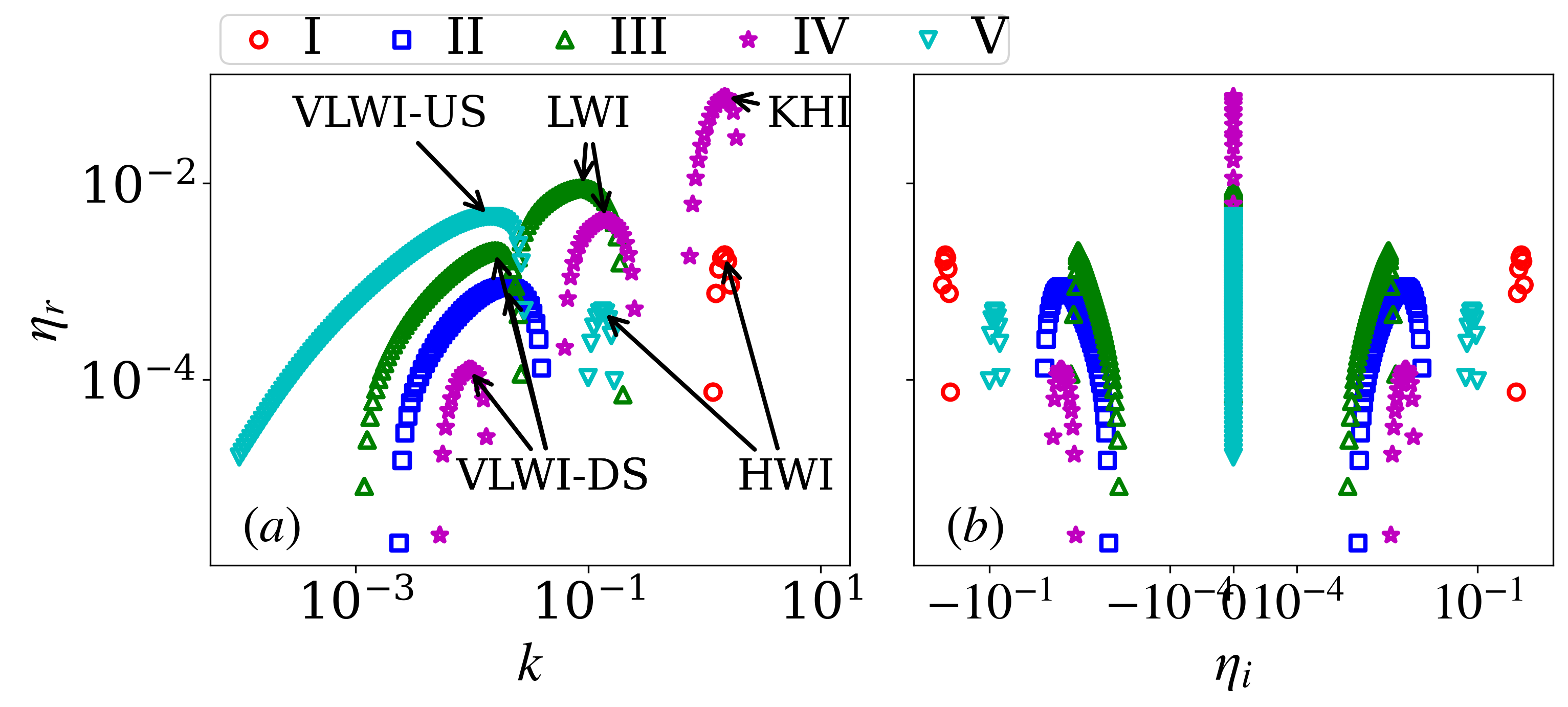}

	\caption{Dispersion relations for typical cases: (a) positive growth rate $\eta_r$ versus wave number $k$ and (b) positive growth rate $\eta_r$ versus wave frequency $\eta_i$. Markers correspond to cases I, II, III, IV, and V from figure~\ref{fig:Qm_ric_theta}.}
	\label{fig:dispRe}
\end{figure}

Figure~\ref{fig:Qm_ric_theta}(a,c) shows the distribution of the growth rate and wave frequency of the fastest-growing modes in the parameter spaces  ($\theta, Ri_b$) and ($\theta, Q_m$), respectively. Examining the contour lines reveals five distinct families of unstable modes, shown schematically in figure~\ref{fig:Qm_ric_theta}(b,d). To better understand these modes, we show the dispersion relation of five representative cases (marked by the symbols in figure~\ref{fig:Qm_ric_theta}) dominated by the five families of instabilities in figure~\ref{fig:dispRe}. The real $\eta_r$ and imaginary $\eta_i$ components of the eigenvalues denote the growth rate and wave frequency (and phase speed $c=-\eta_i/k$) of the unstable mode, respectively. Notably, two of these unstable modes, namely the Holmboe wave instability (HWI) and Kelvin-Helmholtz instability (KHI), can be triggered without the presence of a slope ($\theta=0$, see vertical dotted black line). The other three families of modes rely on the presence of a slope ($\theta\neq 0$) and are named long-wave instability (LWI), \textit{downslope} very-long-wave instability (VLWI-DS), and \textit{upslope} very-long-wave instability (VLWI-US) based on their longer wavelengths ($O(10\sim 10^{4})$) compared to the `short' HWI and KHI ($O(10^{-1}\sim 10)$). To the best of our knowledge, these unstable modes have not previously been investigated in the literature.


We find that the features of these instabilities are generally insensitive to the shapes of base profile and boundary conditions, despite adopting a base profile~(\ref{eq:Us_full}) and no-slip boundary in this section. To support this, we show in \cref{sec:LSA_freeslip} that these instabilities are found using a $\tanh$-shape base state and free-slip boundary condition, as used by \citet{smyth2003turbulence}. This suggests that these instabilities can exist in a wide range of stratified exchange flows along a slope. In the following sections, we characterise the five families of unstable modes in more detail.

\subsubsection{Holmboe wave instability (HWI)}\label{sec:hwi}

The HWI~\citep{holmboe1962behavior} occurs when the density interface is thinner than the shear layer and results from the resonance between vorticity waves at the edges of the shear layer and internal gravity waves at the density interface~\citep{caulfield1994multiple,carpenter2010holmboe}. It gives rise to a pair of counter-propagating growing modes on either side of the density interface.

In SIC, the regime dominated by HWI exists from $\theta=-10^\circ$ to $2^\circ$ and $Ri_b=0.3$ to $4$ ($Q_m=0.1$ to $0.3$) in figure~\ref{fig:Qm_ric_theta}. The dispersion relation of HWI is shown in figure~\ref{fig:dispRe}, where HWI has a pair of complex conjugate eigenvalues with non-zero phase speed $c=-\eta_i/k$.
Despite the well-known feature that HWI can exist in horizontal flows at $Ri_b$ values significantly higher than  $0.25$~\citep{miles1961stability,howard1961note}, we notice that HWI can also be induced over a wide range of $\theta$. More interestingly, the HWI-dominated regime gradually shrinks from $\theta<0$ to $\theta \approx 2$, beyond which HWI ceases to exist. This indicates that increasing downward slopes have a negative effect on HWI, a phenomenon that has not been previously discussed in the literature and constitutes a new result.

\subsubsection{Kelvin–Helmholtz instability (KHI)}\label{sec:khi}

The KHI arises due to the interaction of vorticity waves at two edges of finite shear layers, leading to a sequence of stationary vortex billows that roll up the denser fluids and cause significant mixing~\citep{hazel1972numerical,Smyth1988}. However, unlike these previous studies (with the exception of the recent \cite{ZhuAtoufi2022_II}) the KHI observed here in the SIC geometry is bounded by no-slip solid boundaries at $z=\pm1$. 

In SIC, KHI has a zero phase speed and a characteristic wavelength of $\pi$, consistent with previous studies by~\citet{smyth2019instability,caulfield2021layering,smyth_peltier_1991}. KHI dominates the flow at small $Ri_b\lesssim0.25$, in agreement with the Miles-Howard criterion. Interestingly, like HWI, the longitudinal gravity force can affect the regimes of KHI. The upper bound of the KHI-dominant regime in figure~\ref{fig:Qm_ric_theta}(a) increases linearly from $Ri_b=0.15$ to $0.25$ as $\theta$ increases from $-10$ to $10$. This suggests an enhancement of KHI by a downward slope, which we believe to be an additional new result.

\subsubsection{Long-wave instability (LWI)}\label{sec:lwi}

Of the three new instabilities that arise with slopes, the novel long-wave instability (LWI)  dominates the flow at large downward slopes ($\theta>4^\circ$) and a weak shear (strong stratification). In contrast to KHI and HWI, the LWI has a longer wavelength ($O(10-10^{2})$). Note that the LWI discussed in this paper is distinct from the long waves supported by shallow-water (hydraulic) theory~\citep{lawrence1990hydraulics,ZhuAtoufi2022_II} which are essentially KH waves with a large $k$ (satisfying the hydrostatic approximation) and which can exist at $\theta=0$. LWI, on the other hand, specifically requires $\theta \neq 0$. As depicted in figure~\ref{fig:dispRe}, its phase speed is near-zero.
This instability can be triggered at $Ri_b\gg 1$, at which the shear-induced HWI and KHI vanish. Note that the presence of a mean shear can affect LWI by modifying its growth rate and phase speed. In terms of wave interaction, since vorticity waves vanish as $Q_m\rightarrow0$, we hypothesise that LWI is a result of the interaction between two gravity waves at the density interface whose symmetry is broken by the non-zero slope. However, the $Q_m=0$ condition may be arbitrary when subjected to a non-zero slope, as it requires the gravity and pressure forces to be precisely cancelled by external body forces $\mathcal{F}$ in~(\ref{eq:Mot}). In practice, such a precisely balanced condition is expected to be rarely observed.

\subsubsection{Downslope very-long-wave instability (VLWI-DS)}\label{sec:VLWI-DS}

The new VLWI-DS shares similarities with the LWI, in that it can exist at weak shear (strong stratification) and has a long wavelength. 
However, VLWI-DS  dominates the flow under different conditions, namely when $2^\circ<\theta<5^\circ$ and $Ri_b>0.25$ ($Q_m<0.5$). It is also characterized by very long wavelengths of $O(10^{2}-10^{3})$ (wave numbers $k=O(10^{-3})\sim O(10^{-2})$) and, interestingly, a pair of eigenmodes with complex conjugate phase speeds (figure~\ref{fig:dispRe}). As with the HWI, we thus expect a pair of unstable VLWI-DS modes propagating with opposite phase speeds. The evolution of these unstable long waves and their connections to the onset of turbulence will be further discussed in \S\ref{sec:dns}.

\subsubsection{Upslope very-long-wave instability (VLWI-US)}\label{sec:VLWI-US}

At a negative inclination angle ($\theta<0$), i.e. for upward slopes, another type of very-long-wave instability (VLWI-US) appears with wavelengths $\geq 10^2$ (wave numbers $k<O(10^{-2})$) and a zero phase speed (figure~\ref{fig:dispRe}). This instability is similar to LWI and VLWI-DS in that it requires a slope ($\theta\neq 0$) and can exist in a strongly stratified environment. Contrary to the usually significantly smaller growth rate of the long waves compared with the corresponding short waves, the VLWI-US has in fact a comparable growth rate as HWI; this will be further discussed in \S\ref{sec:LSA_class}.\\

Importantly, these long-wave instabilities have the potential to trigger and sustain turbulence in strongly-stable stratified flows, which are {\em a priori} regarded as stable. 
In \S\ref{sec:dns} we will show that these new instabilities can indeed destabilise the flow at $Ri_b
\gg 1$, eventually resulting in nonlinear bursting and a transition to turbulence and mixing.
It is also important to note that figure~\ref{fig:Qm_ric_theta} only shows the fastest growing modes, whereas multiple families of instabilities can coexist in certain regions, as shown in figure~\ref{fig:dispRe}. As a result, the regions of instability overlap, and the neutral boundary of each instability cannot be identified from figure~\ref{fig:Qm_ric_theta}. In \S\ref{sec:LSA_class}, we will address this challenge by introducing an unsupervised clustering technique to isolate the neutral boundary of each family.
Furthermore, in figure~\ref{fig:Qm_ric_theta}, we include a black line computed from $\gamma=0$, i.e. the natural convective `Thorpe' base state with forcing $\mathcal{F}=0$. Under the  parameters discussed so far ($\Rey=1000$, $\Pran=7$), this line does not overlap with the regimes of long-wave instabilities in  parameter space. 
Nonetheless, it is important to note that different $\Rey$ and $\Pran$ or boundary conditions can modify the regimes of the long wavelength instability and interact with the base flow. An example is demonstrated in \S\ref{sec:re_pr} for $\Pran=28$.

\subsection{Eigenfunctions}\label{sec:eigfun}

\begin{figure}
	\centering		
	\includegraphics[width=.99\linewidth, trim=0mm 0mm 0mm 0mm, clip]{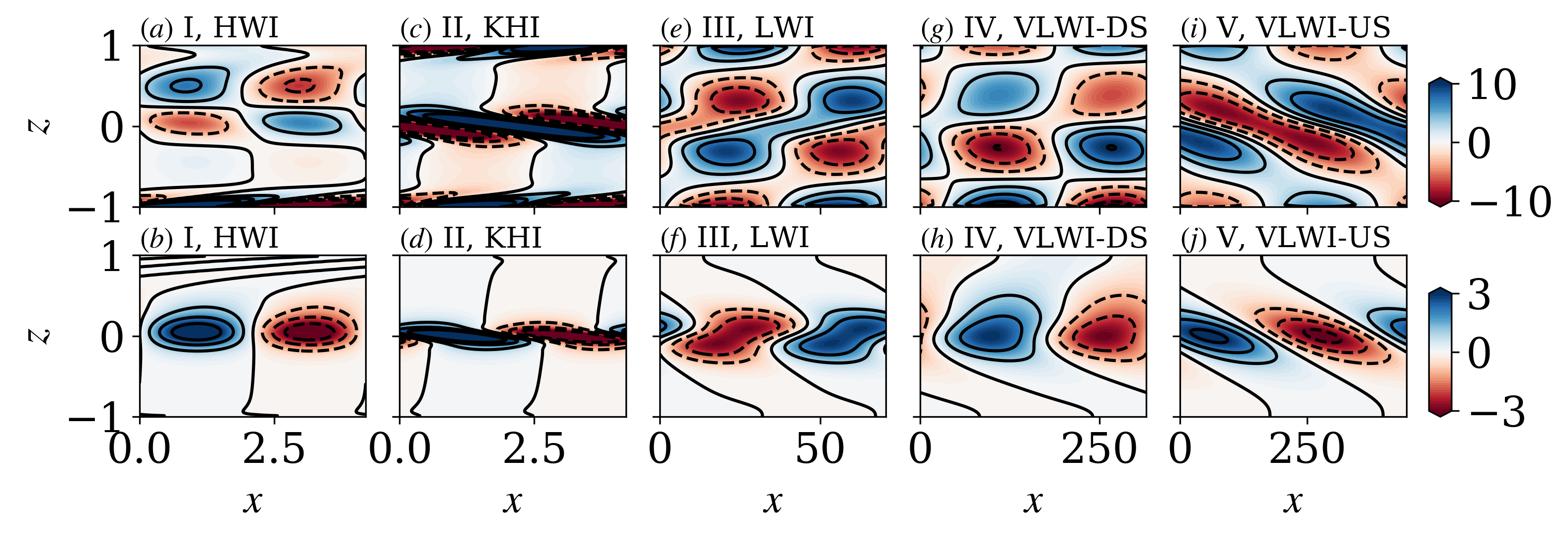}
    \caption{Eigenfunctions of the fastest growing modes for the 5 cases given in figure~\ref{fig:Qm_ric_theta} table~\ref{tab:resol_DNS}: (a-b) I, HWI; (d-f) II, KHI; (g-i) III, LWI; (j-l) IV, VLWI-DS; and (m-o) V, VLWI-US. First row: vorticity eigenfunctions; Second row: density eigenfunctions. }
	\label{fig:eigvec_case}
\end{figure}

Further insights into these SIC instabilities can be gained by examining their eigenfunctions expressed in \eqref{eq:perturbation-form} for representative cases (see figure~\ref{fig:Qm_ric_theta} and table~\ref{tab:resol_DNS}). In figure~\ref{fig:eigvec_case}, we present the vorticity (first row) and density (second row) eigenfunctions of the fastest growing modes for cases I, ..., V, marked in figure~\ref{fig:Qm_ric_theta}, each of which represents one of the five branches of instabilities: HWI, KHI, LWI, VLWI-DS, and VLWI-US, respectively. Note that the $x$-axis in these cases has been re-scaled to compare modes having very different wavelengths. In figure~\ref{fig:eigvec_case}, the wavelengths of HWI and KHI are $\approx 4$, LWI is $\approx 70$, VLWI-DS is $\approx 300$, and VLWI-US is $\approx 420$.

The density eigenfunctions of all modes are concentrated near the interface, indicating the critical role of stratification. Near the walls, the intensity of vorticity eigenfunctions is large due to the no-slip effects of the walls. (Note that, with a free-slip velocity boundary condition, the corresponding modes do not exhibit this intense vorticity at the wall, see \cref{sec:LSA_freeslip}). In the shear layer, one of the HWI modes plotted here (left-propagating) exhibits two pairs of counter-rotating roll cells centred at $z\approx 0.5$. For KHI, the vorticity and density eigenfunctions are highly concentrated at the interface, leaving a weaker bulk region in the rest of channel. By contrast, the vorticity eigenfunctions of LWI, VLWI-DS, and VLWI-US fill the channel and are asymmetric with respect to $z=0$.

\subsection{Neutral boundaries of instabilities}\label{sec:LSA_class}

\begin{figure}
	\centering		
	\includegraphics[width=0.9\linewidth, trim=0mm 0mm 0mm 0mm, clip]{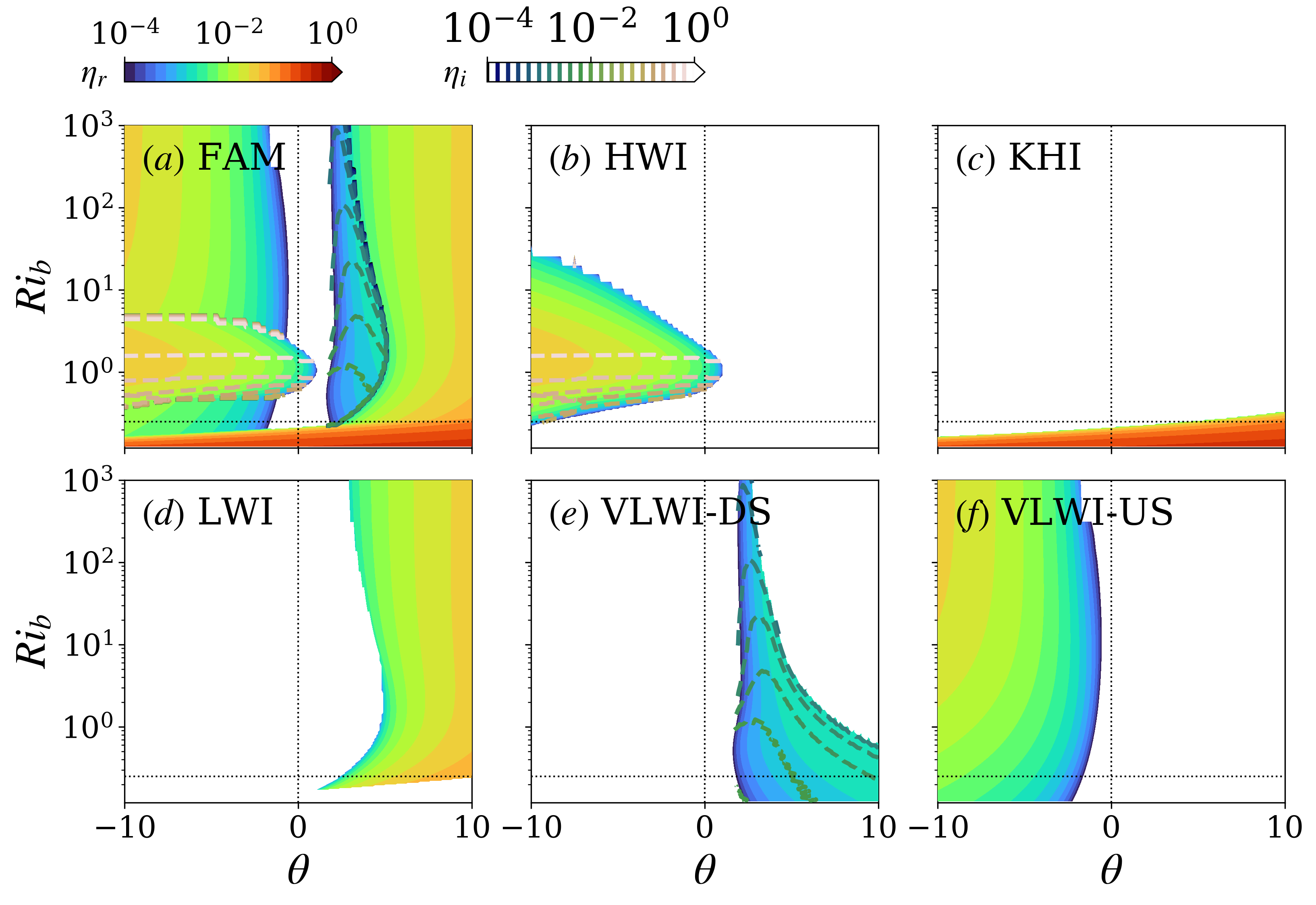}

	\caption{Clustering results: fastest growing mode of each family in $Ri_b-\theta$ parameter space (a) fastest amplifying modes (FAM) of all families reproduced from figure~\ref{fig:Qm_ric_theta}(a), (b) VLWI-US and LWI, (c) HWI modes, (d) KHI modes, and (e) VLWI-DS modes.}
	\label{fig:Ric_theta_class}
\end{figure}

As mentioned in \S\ref{sec:param_space}, different families of instabilities can coexist at the same parameters, making it difficult to determine the neutral boundary of each family from the distribution of fastest growing modes in figure~\ref{fig:Qm_ric_theta}.
To identify the different neutral boundaries we employ an unsupervised machine learning algorithm called DBSCAN (density-based spatial clustering of applications with noise)~\citep{ester1996density}.  The DBSCAN algorithm clusters the local maxima of the dispersion relation (figure~\ref{fig:dispRe}(a)) of all the cases in figure~\ref{fig:Qm_ric_theta} using $k$, $\eta_r$, and $\eta_i$ as input variables. These variables are first logarithmically transformed and normalised before being fed into DBSCAN for clustering. 
Note that the DBSCAN groups the local optimal modes of LWI and VLWI-US together in a single cluster due to their similarity in $k$, $\eta_r$, and $\eta_i$. 
An additional step is taken to distinguish between the two branches by using the fact that LWI occurs when $\theta>0$, while VLWI-US occurs when $\theta<0$.

The clustering analysis in figure~\ref{fig:Ric_theta_class}(b-f) reveals the regimes of different families of instabilities, which could not have been identified by simply looking at the distribution of fastest amplifying modes (figure~\ref{fig:Ric_theta_class}(a)). 
The KHI regime (panel (c)) exactly matches the distribution of the fastest amplifying modes (panel (a)), while other modes (LWI, VLWI-DS, VLWI-US) that overlap with KHI are omitted. This suggests that KHI always has the fastest growth rate. For HWI (panel (b)), increasing $\theta$ clearly decreases the growth rate while shrinking its `territory', causing it to disappear when $\theta>2^\circ$. When $\theta$ is fixed, the fastest growing HWI appears at $Ri_b\approx 1$, while the growth rate decreases as $Ri_b$ departs from $1$. The territory of HWI overlaps with VLWI-US (panel (f)) which can exist when $\theta<-0.5^\circ$. The growth rate of these two modes is comparable so that figure~\ref{fig:Ric_theta_class}(a) cannot display the neutral boundaries of these two modes properly. As for LWI (panel (d)), it generally persists at large positive $\theta$ except for $Ri_b\lesssim 0.2$. The critical $\theta$ for the appearance of VLWI-DS (panel (e)) is $\approx 2.5^\circ$. It overlaps with KHI and LWI at large $\theta$ and small $Ri_b$, respectively, but is mostly omitted in the plot of the fastest growing mode due to its relatively small growth rate.

In general, these long-wave families of instabilities can persist across a wide range of $Ri_b$, ranging from $Ri_b\ll 0.25$ (especially for VLWI-DS and VLWI-US) to $Ri_b\gg 1$. Consequently, we anticipate their widespread presence in sloping stratified exchange flows.

\subsection{Effect of Reynolds and Prandtl numbers}\label{sec:re_pr}

In this section, we study the impacts of Re and Pr on these different families of instabilities. 

\subsubsection{Reynolds number effects}

\begin{figure}
	\centering		
	\includegraphics[width=.65\linewidth, trim=0mm 0mm 0mm 0mm, clip]{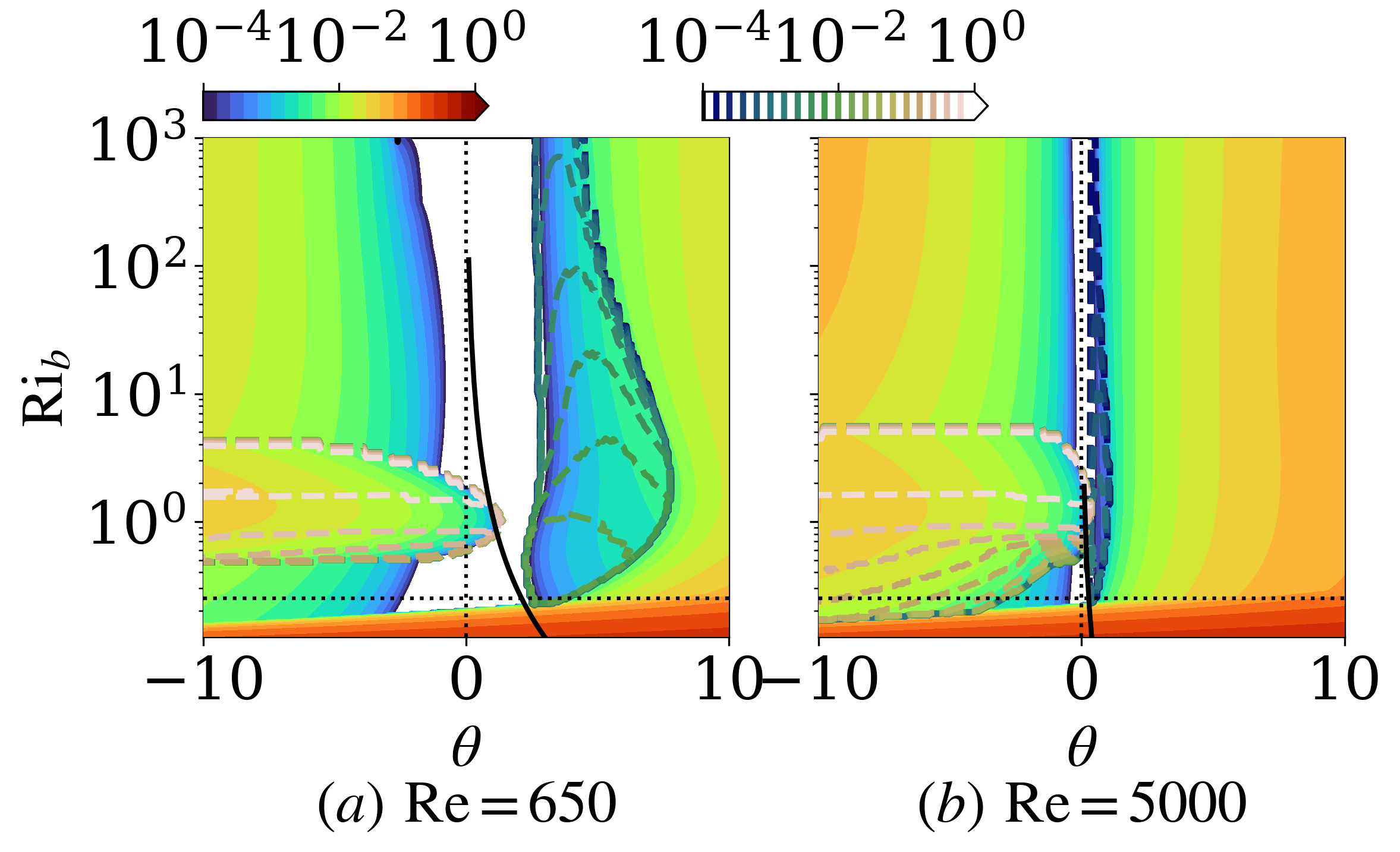}

	\caption{Effect of Reynolds number:  fastest growing mode projected onto  $\mathrm{Ri_b}-\theta$ space for (a) $\Rey=650$, and (b) $\Rey=5000$.}
	\label{fig:Qm_theta_re}
\end{figure}

Figure~\ref{fig:Qm_theta_re} shows the $Ri_b-\theta$ parameter space of the fastest growing modes at a lower $\Rey=650$ (panel (a)) and higher $\Rey=5000$ (panel (b))  than the standard case discussed in \S\ref{sec:param_space}. Generally, $\Rey$ has a significant effect on all families of instabilities except KHI.
The HWI-dominated regime expands to smaller (and slightly larger) $Ri_b$ but shrinks in $\theta$ with increasing $\Rey$. The largest $\theta$ for HWI decreases from $1.3$ to $0.4$, indicating a stronger suppression effect by the slope.
The long-wave families (LWI, VLWI-DS, and VLWI-US) still dominate the large $Ri_b$ region, and their boundaries approach $\theta=0^\circ$ as $\Rey$ increases. For instance, the left-most VLWI appears at $\theta\approx 2$ for $\Rey=650$, whereas it is $\theta\approx0.3^\circ$ for $\Rey=5000$. Similarly, for VLWI-US, the right-most points change from $\theta-=0.6^\circ$ at $\Rey=650$ to $\theta=-0.1^\circ$ at $\Rey=5000$. It is anticipated that in the inviscid limit $\Rey\rightarrow \infty$ the critical $\theta$ will approach $0^\circ$.
Therefore, it is a reasonable speculation that these gravity-induced long waves may be generic in high-$\Rey$ natural water bodies subjected to shear, stratification and even the most shallow slope.

\begin{figure}
	\centering		
	\includegraphics[width=.9\linewidth, trim=0mm 0mm 0mm 0mm, clip]{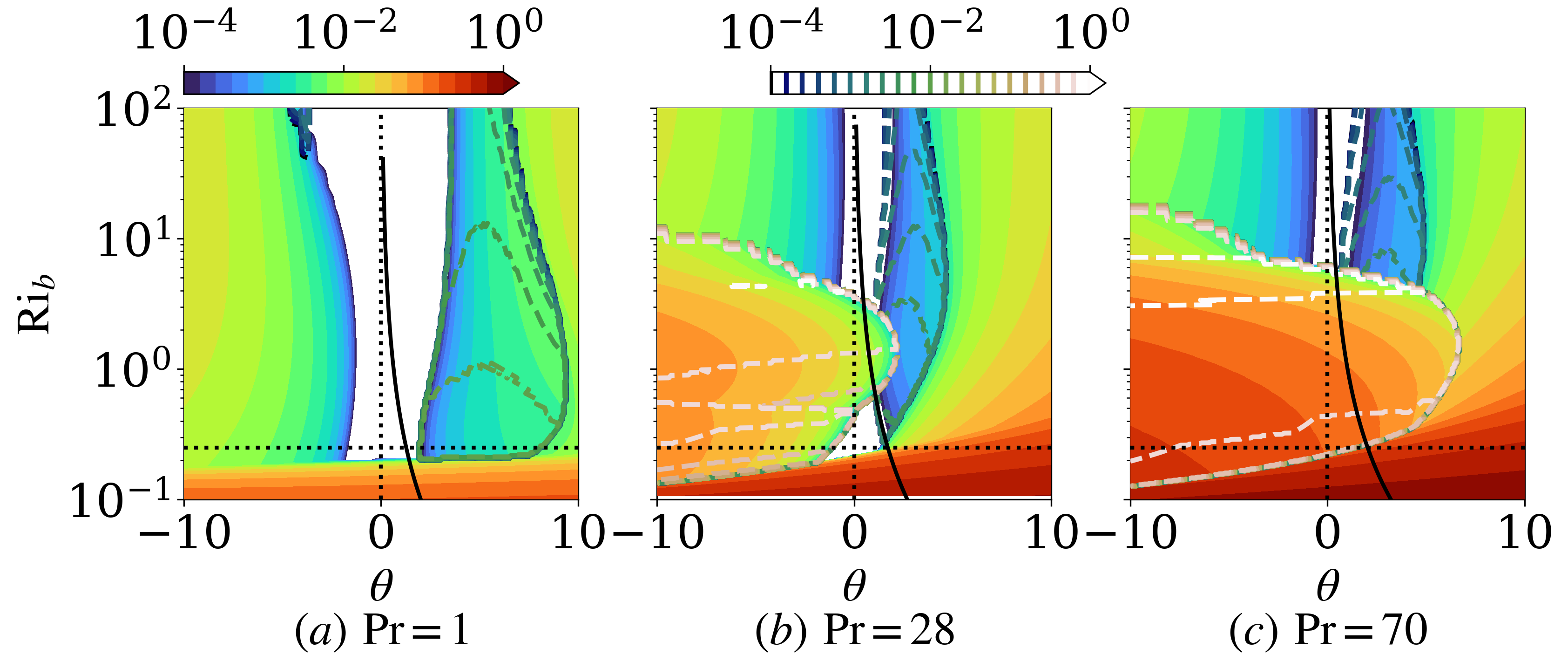}

	\caption{Effect of Prandtl number: fastest growing mode projected onto the $\mathrm{Ri}_b-\theta$ space for (a) $\Pran=1$, (b) $\Pran=28$, and (c) $\Pran=70$, respectively.}
	\label{fig:Qm_theta_pr}
\end{figure}

\subsubsection{Prandtl number effects}

\Cref{fig:Qm_theta_pr} displays the $Ri_b-\theta$ parameter space of the fastest growing modes at $\Pran=1$, $28$, and $70$, respectively, corresponding to the increasingly sharper interface of the density base state, following \eqref{eq:LSA_R}. To ensure convergence, the grid resolution for the LSA was set to $150$, $250$, and $400$, respectively.
As $\Pran$ increases, the influence of the slope $\theta$ on KHI becomes more significant, resulting in a wider upper boundary of KHI, which can be triggered at $Ri_b>0.25$ for large downward slopes $\theta\gtrsim 5^\circ$.
Meanwhile, HWI is also significantly affected by $\Pran$. At $\Pran=1$, HWI does not appear due to the thick density interface determined by (\ref{eq:LSA_R}). However, as $\Pran$ increases, the region of HWI expands significantly towards larger $\theta$.
The long-wave families exist at all $\Pran$.
As $\Pran$ increases from $\Pran=1$ to $28$, the territory of the long waves converges towards $\theta=0$. However, the changes in the territory become less significant from $\Pran=28$ to $70$, indicating a potential convergence of the wave regime at moderate $\Pran$.
However, due to the dominance of HWI at high $\Pran$, the long-wave families are largely omitted by the fastest growing HWI at $Ri_b\lesssim 10$ in figure~\ref{fig:Qm_theta_pr}(c).
Interestingly, at $\Pran=28$, the profile of the Thorpe exchange flow ($\mathcal{F}=0$ in (\ref{eq:Us_full})) passes sequentially through the HWI, VLWI-DS, LWI, and KHI dominated regimes. This provides an example where VLWI-DS and LWI can dominate Thorpe's SIC flow.

\section{Nonlinear evolution of unstable modes}\label{sec:dns}

To gain insight into the subsequent nonlinear evolution of these 
unstable modes we conduct forced two-dimensional direct numerical simulations (DNS).
We describe our DNS in \S\ref{sec:dns_formula} and discuss the evolution and breakdown of the unstable flows in \S\ref{sec:temp_evol}. The instantaneous flow kinetics of these unstable waves and the mechanisms leading to their breakdown are discussed in \S\ref{sec:inst} and \S\ref{sec:mech}, respectively.

\subsection{Forced DNS formulation}\label{sec:dns_formula}

To simulate the growth of linear unstable perturbations on the desired base state, we add to the right-hand sides of  (\ref{eq:ns_mom}) and  (\ref{eq:ns_den}) the two forcing terms 
\begin{equation}
    \label{eq:forc_u_rho}
    F_v = -\frac{1}{\Rey}\frac{\partial^2 U}{\partial z^2}-Ri\sin\theta R,   \qquad
    F_\rho = -\frac{1}{\Rey\Pran}\frac{\partial^2 R}{\partial z^2},   
\end{equation}
respectively. 
In this way, the mean velocity and density of the DNS are forced towards the targeted base profile of $U(z)$ and $R(z)$.
These terms can be regarded as enforcing a pressure-driven exchange flow under a sustained stratification. Similar approaches that apply body forces to the stratified flows were introduced in ~\citet{taylor2016new} and \citet{smith2021turbulence}.

We perform the simulations using the open-source solver Dedalus~\citep{burns2020dedalus} employing a Fourier-Chebyshev pseudo-spectral scheme for spatial discretisation and a 3rd-order, 4-stage diagonally-implicit+explicit Runge-Kutta scheme~\citep{ascher1997implicit} for time stepping. We imposed periodic boundary conditions in the streamwise $x$ direction, while we applied no-slip and no-flux boundary conditions for velocity and density, respectively, to the solid walls at $z=\pm 1$, as in the LSA. The streamwise length $L_x$ of the channel was set equal to the wavelength of the fastest growing mode, while the channel height $L_z$ was fixed at $2$. We employed a uniform grid for the $x$ direction and a Chebyshev grid for the $z$ direction. The simulation resolution was determined by the geometrical and physical parameters of the problem. 
We initialised the simulations by superimposing on the base state the eigenfunctions of the LSA unstable modes with a perturbation magnitude $\zeta$.
The parameters of the production runs are listed in \cref{tab:resol_DNS}.

    \begin{table}
        \centering
        \begin{tabular}{lcccccccccc}
            \hline
             Instability & Case & $\Rey$ & $\Pran$ & $\kappa$ & $\theta$ (deg.) & $\gamma$ & $k$ & $\zeta$ & $L_x$ & $N_x\times N_z$  \\ 
             \hline
             HWI & I & \multirow{4}{*}{1000} & \multirow{4}{*}{7} & \multirow{4}{*}{2} & 0.5 & -0.0028 & 1.5 & 0.2 & 4.2 & $320\times 144$ \\
             KHI & II & & & & 2 & 0.0077 & 1.5 & $10^{-6}$ & 4.2 & $320\times 144$ \\
             LWI & III & & & & 6 & 0.093 & 0.089 & 0.01 & 70.5 & $960\times 144$ \\ 
             VLWI-DS & IV & & & & 3 & 0.041 & 0.02 & 0.5 & 314.9 & $6000\times 144$ \\ 
             VLWI-US & V & & & & -6 & -0.125 & 0.02 & 0.5 & 444.7 & $6000\times 144$ \\ 
            \hline
        \end{tabular}
        \caption{Numerical parameter values used for the DNS  runs.}
        \label{tab:resol_DNS}
    \end{table}

\subsection{Temporal evolution}\label{sec:temp_evol}

\begin{figure}
	\centering		
	\includegraphics[width=.7\linewidth, trim=0mm 0mm 0mm 0mm, clip]{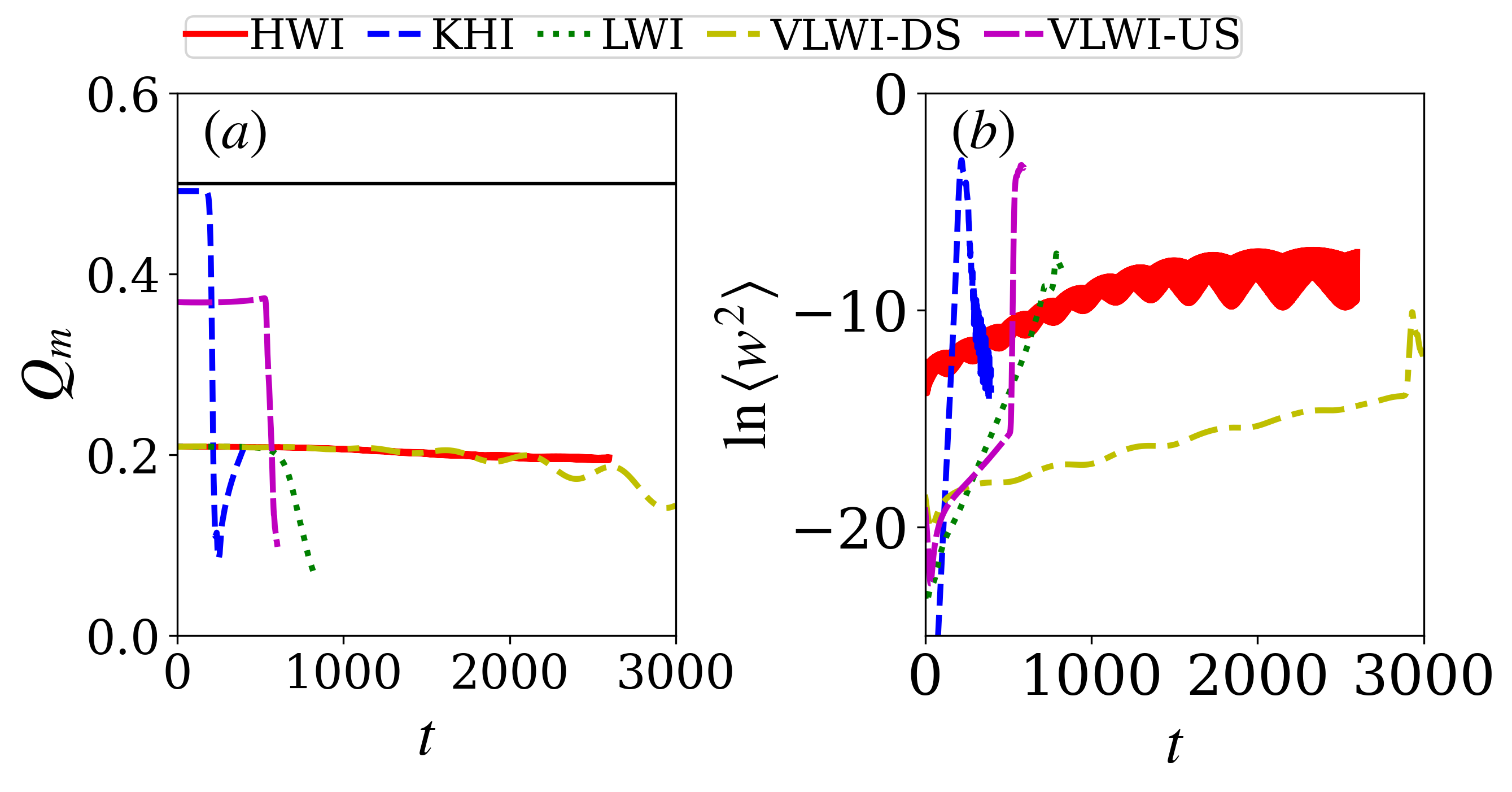}

	\caption{Time evolution of (a) mass flux $Q_m$ and (b) logarithm of vorticity squared $\ln\langle w^2\rangle$ for the fastest growing modes in table~\ref{tab:resol_DNS}. The slopes of the growth for the HWI, KHI, LWI, VLWI-DS, and VLWI-US modes are $0.0038$, $0.15$, $0.018$, $0.0017$, and $0.0089$, respectively, consistent with $2\eta_r$ of corresponding unstable mode in LSA ($0.0037$, $0.15$, $0.018$, $0.00018$, and $0.0092$).} 
	\label{fig:forc_ts}

\end{figure}

In this section, we focus on the temporal evolution of the fastest growing modes of each instability family, I, II, III, IV, and V, as marked in figure~\ref{fig:Qm_ric_theta}. 
Figure~\ref{fig:forc_ts} shows the temporal behaviour of the unstable modes through the time series of the mass flux $Q_m(t)$ (\ref{eq:Qm}) and the spatially-averaged vertical velocity of perturbations $\langle w^2 \rangle(t)$, where $\langle \cdot \rangle$ denotes averaging over $x,z$. The magnitude  $\zeta$ of the perturbation was chosen differently for each mode in order to obtain a reasonably long linear growth period. The forcing magnitude $\gamma$ is determined so that the base velocity matches the selected cases in figure~\ref{fig:Qm_ric_theta}. The exchange flow is simulated by forcing the background flow in time using~(\ref{eq:forc_u_rho}) and allowing the perturbations to grow. 

In figure~\ref{fig:forc_ts}(a), the background state is controlled by the body forces (\ref{eq:forc_u_rho}) so that $Q_m$ initially remains constant and consistent with the targeted base state (as marked in figure~\ref{fig:Qm_ric_theta}) until the perturbations are significantly amplified and the flow enters the nonlinear stage.
This initially constant $Q_m$ value indicates the effectiveness of the forcing method to maintain a sustained background state before the intense nonlinear dynamics set in.
The evolution of the disturbance amplitudes is shown in figure~\ref{fig:forc_ts}(b), with all cases exhibiting a clear exponential growth period for $w^2$, with growth rates matching the corresponding linear unstable modes.
For KHI, LWI, and VLWI-US, following the exponential growth period, an intense nonlinear bursting process is caused by the breakdown of the primary waves, leading to intense mixing and changes in $Q_m$ and $w^2$. 
In contrast, the sudden changes in $Q_m$ do not appear for HWI and VLWI-DS since their primary waves do not break down.
HWI and VLWI-DS have a pair of conjugate modes, represented by oscillating $w^2$ profiles, due to the synchronization of complex-conjugate modes, as discussed in \cite{yang2022velocity}. 
Interestingly, after the nonlinear bursting at $t=1250$, the nonlinear HWI still maintains the oscillating pattern~\citep{lefauve2018structure}.
The time series of $Q_m$ and $\ln\langle w^2\rangle$ pinpoint the critical time when the nonlinear effects become prominent. Specifically, this occurs when $Q_m$ deviates from its constant level or when $\ln\langle w^2\rangle$ no longer shows exponential growth after reaching a certain amplitude.
Note that the critical amplitude for nonlinear bursting remains independent of the initial amplitude of perturbations. However, it varies for each individual unstable mode, as illustrated in figure~\ref{fig:forc_ts}.
The critical time may vary depending on the particular unstable mode, the growth rate, and the magnitude of the initial perturbation.

\begin{figure}
	\centering		
	\includegraphics[width=.99\linewidth, trim=0mm 0mm 0mm 0mm, clip]{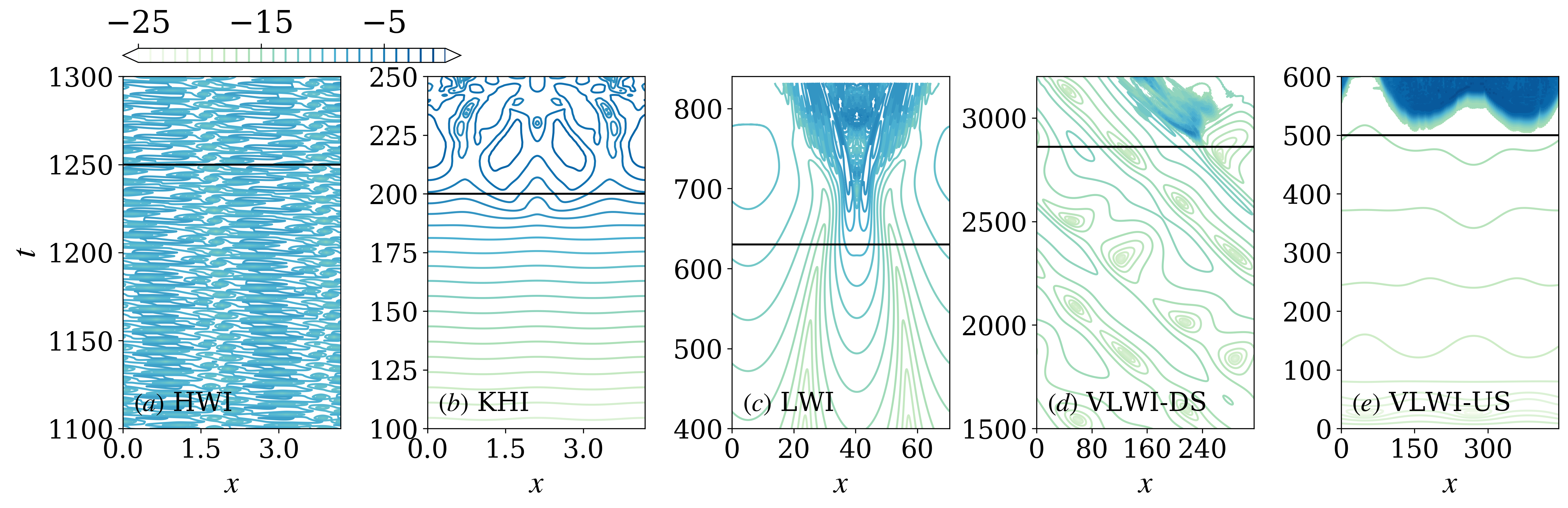}

	\caption{Spatial-temporal diagrams of $\ln\langle w^2\rangle_{z}$ from the nonlinear simulations of (a) HWI, (b) KHI, (c) LWI, (d) VLWI-DS, and (e) VLWI-US. The black solid lines indicate the times of visibly nonlinear dynamics identified in figure~\ref{fig:forc_ts}.}	\label{fig:forc_w2_xt}
\end{figure}

Figure~\ref{fig:forc_w2_xt} shows the $x-t$ diagrams of $\ln\langle w^2\rangle_{z}$, where $\langle \cdot \rangle_{z}$ indicates $z-$averages.
In case HWI (panel (a)), we observe left-going waves from the spatial-temporal diagram, while its conjugate pair is omitted as only one mode of the pair is imposed as initial perturbation in the DNS. Nonlinear effects become significant at $t\approx 1250$ ($\ln{\langle w^2 \rangle}\approx -9$) owning to a relatively small growth rate ($0.0038$), as indicated by the saturation of the exponential growth of $w^2$ in figure~\ref{fig:forc_ts}(b). Interestingly, the spatial-temporal pattern of HWI does not change significantly after $t\approx 1250$ in figure~\ref{fig:forc_w2_xt}(a). This implies that nonlinear effects only halt the linear growth of HW structures, which maintain their forms as nonlinear HW, as observed in experiments and nature~\citep{meyer2014stratified,cudby2021weakly}.
By contrast, KHI generates strong secondary instabilities~\citep{mashayek_peltier_2012} after the onset of nonlinear effects at $t\approx 200$ ($\ln{\langle w^2 \rangle}\approx -5$). Consequently, the KH billows break up, leading to highly chaotic flow stages with small-scale structures.

In figure~\ref{fig:forc_w2_xt}(c-e), the $x-t$ diagrams of $\ln\langle w^2\rangle_{z}$ reveal that for the three new families of long waves, small-scale structures emerge in the latter stages of the transitions, characterised by highly fluctuating contour lines. In the case of LWI (figure~\ref{fig:forc_w2_xt}(c)), the onset of an intense chaotic flow period occurs at $t=710$, during which small-scale structures are initially generated at $x=25$ and $x=55$ where $\ln\langle w^2\rangle_{z}$ of the linear wave peaks. These structures then propagate towards the quiet regions and ultimately trigger a disorganised flow field across the channel. Interestingly, we have observed from figure~\ref{fig:forc_ts}(a) that the nonlinear effects set in at $t=630$ ($\ln{\langle w^2 \rangle}\approx -14$) as $Q_m$ significantly deviates from the original constant level.
At this stage, the two peaks of the linear disturbance approach each other while contour curves twist. The $\ln\langle w^2\rangle_{z}$ distribution no longer maintains its shape as is in the linear growing period. As we will show later, this nonlinear dynamics is the breakdown of the long waves.

In case VLWI-DS, as shown in figure~\ref{fig:forc_w2_xt}(d), the unstable wave moves leftward and grows exponentially until nonlinear dynamics set in at $t=2860$ ($\ln{\langle w^2 \rangle}\approx -19$), which is identified by a jump in $\langle w^2 \rangle(t)$ in figure~\ref{fig:forc_ts}(b). Small-scale waves/structures are formed at a local peak of the long wave $w^2$, which propagate both leftward and rightward, creating strong mixing. Despite the intense bursting of the flows, the long wave does not break down like LWI, presumably due to its low growth rate. It continues to propagate at the same phase speed as the linear wave energy that was previously used to amplify the long wave and is then fed to the small-scale waves, which eventually break down and dissipate, allowing the long wave to persist for a long period of time and propagate over a long distance.

For VLWI-US (shown in figure~\ref{fig:forc_w2_xt}(e)), local nonlinear bursting and small-scale structures are directly created at $t=500$ ($\ln{\langle w^2 \rangle}\approx -16$) and $x\approx 150$ and $350$ on top of the long wave. Similar to VLWI-DS, a preliminary breakdown of the long-wave is not observed. Soon after, intense secondary instabilities fill the entire channel and the long waves are no longer distinguishable.

In summary, the evolution of these linear long waves eventually leads to the appearance of nonlinear dynamics and intense secondary short-wavelength structures. 
The formation of these small-scale structures is a result of perturbation amplification, which alters the base state and allows the growth of short-wave instabilities. We will explore this mechanism in more detail in \S\ref{sec:inst}-\ref{sec:mech}.

\subsection{Features of flow kinematics}\label{sec:inst}

In this section we present instantaneous flow fields corresponding to the key stages of evolution for the long wave instabilities. The kinematics of the short waves, i.e., HWI and KHI, have been well-documented in the literature~\citep{smyth2003turbulence,salehipour_peltier_mashayek_2015,mashayek_peltier_2012,mashayek_peltier_2013,lefauve2018structure}, and will not be repeated here.

\begin{figure}
	\centering		
	\includegraphics[width=.99\linewidth, trim=0mm 0mm 0mm 0mm, clip]{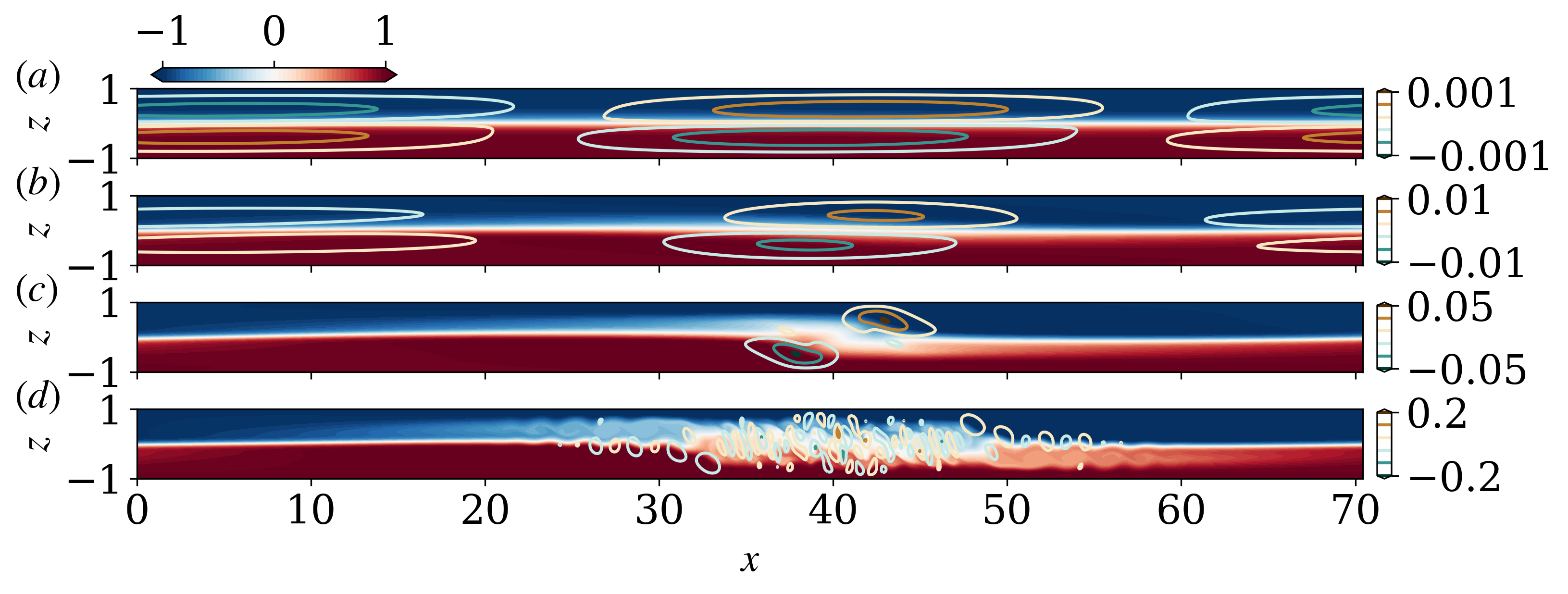}

	\caption{Nonlinear LWI: density (colour) and vertical velocity (lines) snapshots of the forced DNS at $K=0.01$ at different time instances: (a) $t=400$, (b) $t=600$, (c) $t=700$, and (d) $t=760$.}
	\label{fig:forc_III_rho}
\end{figure}

\begin{figure}
	\centering		
	\includegraphics[width=.99\linewidth, trim=0mm 0mm 0mm 0mm, clip]{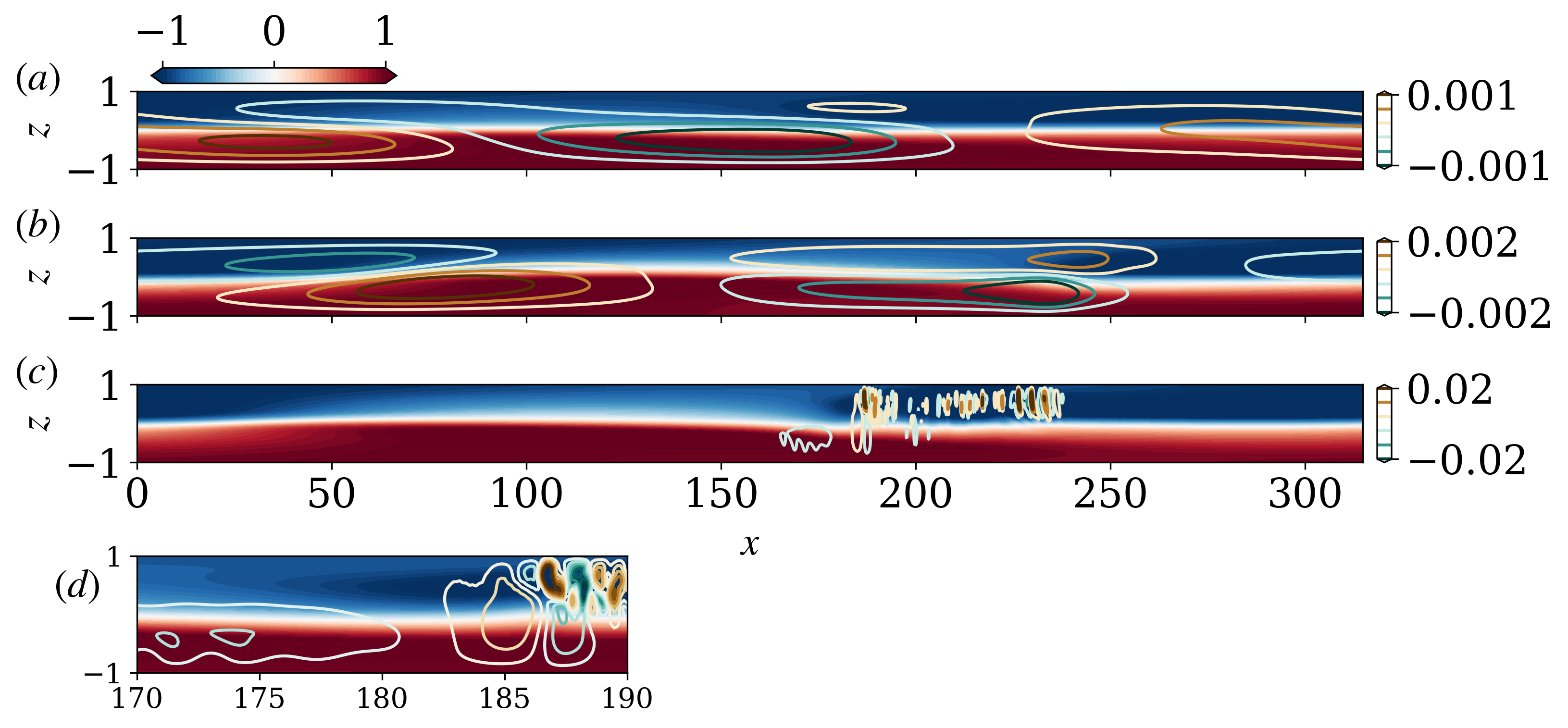}

	\caption{Nonlinear LWI-DS: density (colour) and vertical velocity (lines) snapshots of the forced DNS at $K=0.5$: (a) $t=1500$, (b) $t=2500$, and (c) $t=2900$. An enlarged plot of panel (c) is shown in panel (d).}
	\label{fig:forc_IV_rho}
\end{figure}

\begin{figure}
	\centering		
	\includegraphics[width=.99\linewidth, trim=0mm 0mm 0mm 0mm, clip]{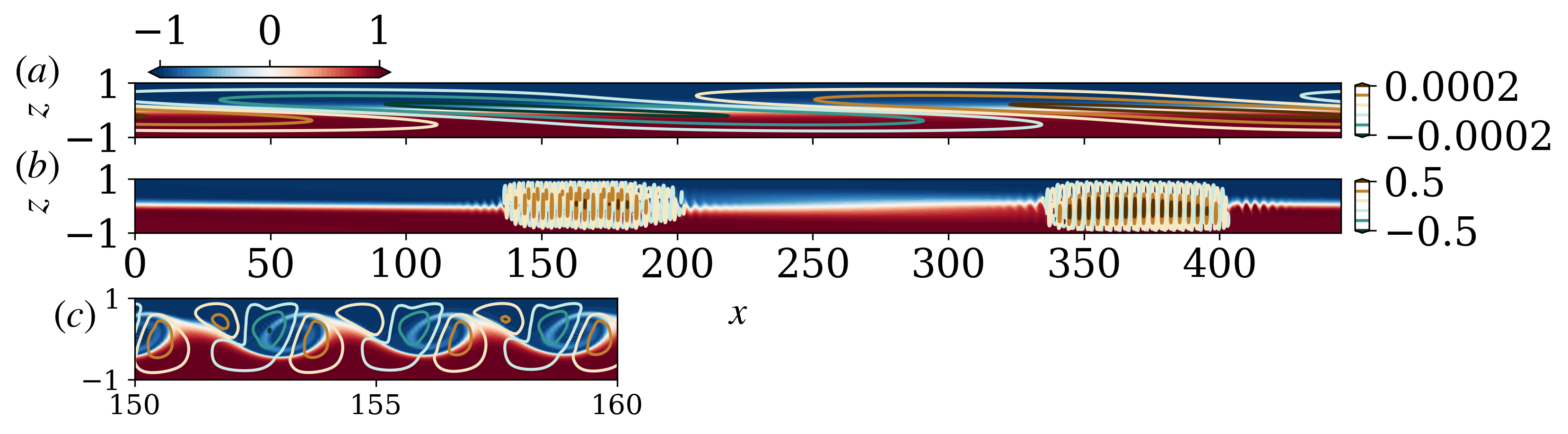}

	\caption{Nonlinear VLWI-US: density (colour) and vertical velocity (lines) snapshots of the forced DNS  at $K=0.1$: (a) $t=600$, and (b) $t=888$. An enlarged plot of panel (b) is shown in panel (c).}
	\label{fig:forc_VLWI2_rho}
\end{figure}

Figure~\ref{fig:forc_III_rho} shows snapshots of the total density $\rho=R+\rho^\prime$ (colours) and vertical velocity $w^\prime$ (lines) at four stages of LWI. We find two stages of nonlinear breakdown corresponding to the breakdown of the long wave and the generation of KH-like overturns. At the linear stage (figure~\ref{fig:forc_III_rho}(a)), the impacts of the density perturbation on the mean are barely observable and the interface is thin and flat. Later, the growth of LWI raise and drop on the left- and right-hand sides of $x=40$, creating a large-scale jump (figure~\ref{fig:forc_III_rho}(b)). Meanwhile, $w^\prime$ becomes localized at $x=40$. Further amplification of the unstable mode breaks down the jump, generating a chaotic region at $t=700$. At this stage, the instability is no longer linear, as demonstrated in \S\ref{sec:temp_evol}. At $t=760$, a series of short overturns resembling to KH billows are formed inside and at two sides of the chaotic region at $x=40$. These waves propagate away from the chaotic region and may eventually lead to (two-dimensional) turbulence.

\Cref{fig:forc_IV_rho} shows snapshots of the flow fields at three stages of VLWI-DS. From $t=1500$ (panel a) to $t=2500$ (panel b), VLWI-DS amplifies, while light-blue (e.g., upper layer: $150<x<200$) and light-red (e.g., bottom layer: $250<x<300$) regions become distinguishable, indicating the enhancement of mixing in these regions which acts to dissipate the  energy injected by gravity. As the base flow is frozen by the simulation, the mixing is attributed to the amplification of VLWI-DS.
At $t=2900$ (panel c), further growth of the long wave induces intense KH-like overturns near the leading edge of VLWI-DS, characterised by the strong fluctuation of $w^\prime$ in the range of $x=175\sim 250$. These overturns create extra dissipation and mixing of the flow, acting to balance the extra kinetic energy supplied by gravity. In contrast to the KH-like overturns in LWI, the overturns can center within the bulk flow of each layer in addition to the interface (figure~\ref{fig:forc_IV_rho}(d)). This is because the propagation of the leading edge of the long waves (dark blue region at $x=180$) into the mixed region (light blue region at $x<180$) creates a weak interface between the denser and lighter regions inside the flow layer.

Finally, in the VLWI-US case (figure~\ref{fig:forc_VLWI2_rho}), the amplification of the stationary waves directly induces localized KH-like billows characterised by strong fluctuations of $w^\prime$ in panel (c) at the interface without first breaking down as in LWI. This behaviour may be due to the slower growth rate, which prevents the formation of a distinct nonlinear `jump' observed in LWI. 

As discussed in \S\ref{sec:temp_evol}, the evolution of these long wave families eventually lead to intense bursting processes that form strong small-scale KH-like overturns. These overturns are responsible for dissipating the kinetic energy injected by a positive slope that cannot be completely balanced by the dissipation of long waves. Note that in all the long waves cases, a short-wave instability (KHI and HWI) does not initially exist according to the LSA in \S\ref{sec:param_space}. In the next section, we study how these short-wave KH-like overturns are induced by nonlinear long waves.

\subsection{Breakdown mechanism}\label{sec:mech}

In \S\ref{sec:param_space}, we showed that the KHI can only occur when $Ri_b\lessapprox 0.25$. In the new long wave cases considered in this study, $Ri_b\gg 1$, hence KHI cannot be triggered. Instead, KH-like overturns are formed by the nonlinear evolution of these long waves.

To understand the cause of the formation of KH-like overturns, we computed the gradient Richardson number $Ri_g$ at the total density interface $\rho=0$, which is defined as
\begin{equation}
Ri_g(x,t) \equiv Ri \frac{\partial \rho/\partial z}{(\partial u/\partial z)^2}\Big|_{\rho=0},
\end{equation}
where we recall that $Ri\equiv 1/4$ (see \S~\ref{sec:gov-eqs}). Note that the field $Ri_g(x,t)$ is based on the total velocity $u$ and density $\rho$, and thus differs from the constant $Ri_b$ defined in \eqref{eq:Ri_b}, which is based on the initial base flow profiles $U$ and $R$.

\begin{figure}
	\centering		
	\includegraphics[width=.8\linewidth, trim=0mm 0mm 0mm 0mm, clip]{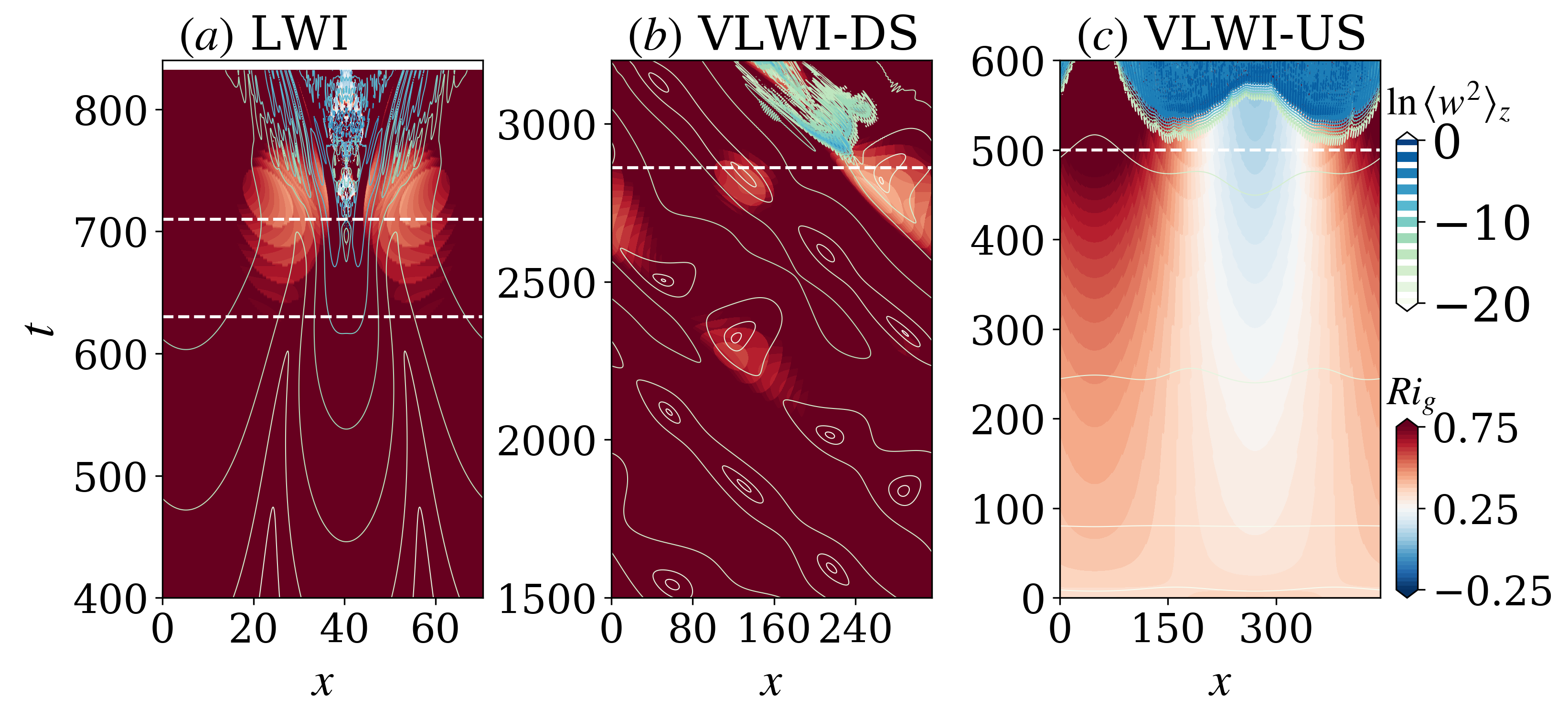}

	\caption{Spatial-temporal diagrams of the gradient Richardson number $Ri_g$ at the density interface in forced DNS for (a) LWI, (b) VLWI-DS, and (c) VLWI-US. The color maps show the values of $Ri_g$, while the lines represent $ln\langle w^2\rangle_{z}$.}
	\label{fig:forc_rig_xt}

\end{figure}

In figure~\ref{fig:forc_rig_xt}, we show the $x-t$ diagrams of  $Ri_g$, with $w^2$ contour lines superimposed. In general, as the unstable long waves grow, the density gradient can decrease due to diffusivity which increases the mixing layer. Meanwhile, the velocity gradient can increase as the perturbations are amplified.
These results in a decreasing interfacial $Ri_g$, until it reaches below $0.25$ (shades of blue), with which small-scale structures are associated.

For each individual long-wave family, the process is slightly different. LWI (figure~\ref{fig:forc_rig_xt}(a)) has two stages of nonlinear breakdown. The first stage appears at $t=680$ when a `jump' is formed at $x=45$. This jump changes the density interface and generates two low $Ri_g$ regions separated by a high $Ri_g$ region. In these low $Ri_g$ regions, $Ri_g$ continuously decreases due to the amplification of long waves and eventually reduces below $0.25$, which potentially allows the growth of the secondary KHI in these regions. Finally, overturns are formed in these regions, leading to the second stage of nonlinear breakdown.
Similarly, the amplification of VLWI-DS (figure~\ref{fig:forc_rig_xt}(b)) also causes a low $Ri_g$ region that travels along with the waves. As soon as $Ri_g<0.25$, intense overturns are formed in this region and later contaminate the entire duct.
For VLWI-US, the overturns are first formed at the edges of the low $Ri_g$ region ($170<x<330$). The close relation between the low $Ri_g$ region and the onset of nonlinear short waves strongly suggests that the overturns are a consequence of the decreasing of local $Ri_g$ caused by the nonlinear evolution of initially long waves.

\begin{figure}
	\centering		
	\includegraphics[width=.9\linewidth, trim=0mm 0mm 0mm 0mm, clip]{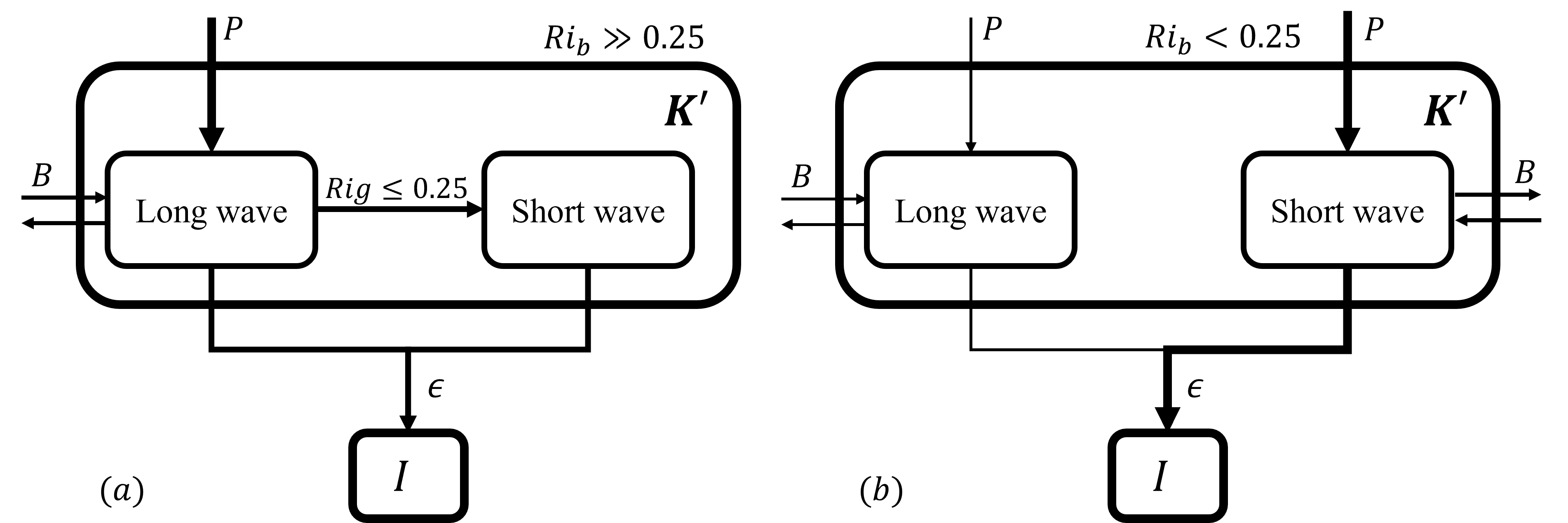}
	\caption{Pathways of turbulent kinetic energy in sloping exchange flows under (a) strong stratification $Ri_b \gg 0.25$, where only very long waves are unstable, (b) weaker stratification $Ri_b< 0.25$ where both short and long waves coexist.}
	\label{fig:tke_budget}

\end{figure}

From an energy budget perspective, the formation of short-wave overturns in these long-wave simulations allows for more efficient dissipation of kinetic energy fed by external forces. In figure~\ref{fig:tke_budget}(a), we illustrate the pathways of turbulent kinetic energy $K^\prime$ as given schematically by
\begin{equation}
\partial_t K^\prime=\Phi^{K^\prime}+P-B-\epsilon,
\end{equation}
where, $P$, $\epsilon$, $B$, and $\Phi^{K^\prime}$  represent the production, dissipation, buoyancy flux, and transport terms of $K^\prime$. The reader may refer to \citep{caulfield2021layering,lefauve_experimental2_2022} for the definition and a more comprehensive discussion of the kinetic budget of stratified shear flows. When $Ri_b\gg 0.25$, initially only the long waves are allowed to grow in the flow with strong stratification, gaining energy from the mean flow through production and buoyancy terms and losing it through dissipation. As the long waves are amplified, local shear is created and amplified by the growing velocity perturbations, leading to the decrease of local $Ri_g$. As $Ri_g\lesssim 0.25$, the necessary condition for the growth of short waves (mostly KHI here) is satisfied. The short waves then grow, extracting $K^\prime$ from the long waves and dissipating it to internal energy. This opens a new energy pathway that allows flows with strong stratification (large $Ri_b$) to dissipate energy by creating small-scale (turbulent) structures.  
When $Ri_b\ll 0.25$ (figure~\ref{fig:tke_budget}(b)), long waves can coexist with short waves (e.g. case IV in figure~\ref{fig:dispRe}) and may contribute to the energy dissipation. But they are often significantly weaker than short waves since short waves tend to have a faster growth rate. Meanwhile, the short wave directly gains most of the kinetic energy from the mean flow and converts it to internal energy. We also note that turbulence created by these unstable waves can also induce irreversible mixing which, in return, contributes to the production of internal energy.

  \section{Conclusions}\label{sec:conclusions}
  
In this paper, we examined the effects of longitudinal gravitational forces on the  stability of two-layer stratified exchange flows by conducting linear stability analyses and nonlinear forced DNS in a sloping channel with solid top and bottom boundaries.
In addition to the well-known Holmboe and Kelvin-Helmholtz instabilities, we revealed the existence of three new families of long-wave instabilities subject to non-zero gravitational forces ($\theta\neq 0$):

\begin{itemize}
\item Long-wave instability (LWI), with wavelengths of the order $10-100$ channel depths (wave number $k=O(10^{-2}-(10^{-1})$) and a near-zero wave speed;
\item Downslope very-long-wave instability (VLWI-DS), with wavelengths of the order $100-1000$ channel depths (wave number $k=O(10^{-3})\sim O(10^{-2})$), a non-zero wave speed, and complex conjugate eigenmodes implying travelling waves;
\item Upslope very-long-wave instability (VLWI-US), with wavelengths $\geqq 100$ channel depths (wave number $k<O(10^{-2})$) and a near-zero wave speed.
\end{itemize}

The LWI and VLWI-DS exist at a positive (favourable) slope where the along-slope component of gravity reinforces the pressure gradient, while VLWI-US emerges under a negative (adverse) slope condition.
Interestingly, their onset is largely independent of the base flow speed. As a result, they can be triggered even  at very high background gradient Richardson numbers $Ri_b\gg 1$,
and induce chaos and sustain (two-dimensional) turbulence and mixing in strongly stratified fluids.
In a weakly stratified flow (low $Ri_b$), they can also coexist with short-wave instabilities (KHI and HWI), but they generally have a lower growth rate.  The short-wave HWI and KHI also exhibit interesting features under non-zero slopes. Increasing $\theta$  tends to suppress the HWI regime while enhancing the KHI. Moreover, the neutral boundary of KHI increases linearly from $Ri_b=0.15$ at $\theta=-10^\circ$ to $0.25$ at $\theta=10^\circ$.

The long-wave families appear under broad flow conditions. To explore their dependence on flow parameters, we varied the Reynolds number $\Rey$, Prandtl number $\Pran$, the base flow and boundary conditions. While increasing the $\Rey$ does not significantly affect KHI, it does enhance the other instabilities. The range of HWI expands to larger $Ri_b$, while the range of the long-wave instabilities approaches $\theta=0$. Therefore, it can be anticipated that as $\Rey\rightarrow \infty$ (as is often the case in natural flows), the critical slope required to trigger these long instabilities approaches zero $|\theta|\rightarrow 0$. 
Increasing $\Pran$, or equivalently, decreasing the thickness of the density interface of the base flow, cause the unstable range of HWI to expand towards larger $\theta$ and smaller $Ri_b$. Meanwhile, the range of the long-wave instabilities slowly approaches $\theta=0$, indicating that these long waves can exist in both water (with $\Pran$ ranging from 7 to 700) and air ($\Pran\approx 1$).
It should be noted that these instabilities are not limited to the sine-like base state and the no-slip boundary conditions used in this study. Instead, they can be triggered by, e.g., a $\tanh$-shaped velocity base and free-slip (but impenetrable) velocity boundary conditions.

Finally, we studied the nonlinear evolution of the different instabilities in the inclined channel and their connections to turbulence using a two-dimensional forced DNS that maintains the base states.
For all of the long-wave instabilities, the evolution eventually led to a nonlinear bursting process with significant small-scale secondary KH-like overturns and mixing.
Specifically,  the LWI exhibit two nonlinear stages  where an initial breakdown of the long waves is followed by a secondary bursting process, creating multiple intense KH-like overturns.
For VLWI-DS and VLWI-US, the long waves do not break down. Instead, they directly alter the base states and induce localized small-scale overturns. 

The evolution of these long instabilities results in a decrease in the density gradient and an increase in the shear, which in turn reduces the local gradient Richardson number $Ri_g$. Our analysis reveals that the appearance of KH-like overturns is highly correlated with a local low $Ri_g$, which approaches the critical threshold of $0.25$ (below which we find the KHI), substantiating the emergence of localised KH-like overturns. From a turbulent kinetic energy budget perspective, a new energy pathway allows the transfer of kinetic energy from the mean flow to the long waves (linearly) and then to the short waves (nonlinearly), eventually leading to the dissipation of turbulent kinetic energy, under conditions where short waves are linearly stable.

The circumstances under which turbulence can persist in strongly stratified flows remains a fascinating debate within the community~\citep{caulfield2021layering}. We demonstrated that weakly unstable (very) long waves may trigger turbulence and mixing after long periods of time, even under initially very strongly stratified conditions ($Ri_b\gg 1$). These results have particular relevance for high$-\Rey$ flows in rivers~\citep{yoshida1998mixing} and straits~\citep{gregg2002flow}, or any natural flow having even very shallow slopes $\theta\approx 0$. 
A quantitative investigation of the turbulent transition and mixing associated with these long waves would require three-dimensional direct numerical simulations, an endeavour left for future work.

\vspace{0.3cm}

\textbf{Acknowledgments}\par
We acknowledge the ERC Research and Innovation Grant No 742480 `Stratified Turbulence And Mixing Processes' (STAMP). A. L. acknowledges a Leverhulme Trust Early Career Fellowship and a NERC Independent Research Fellowship (NE/W008971/1). For the purpose of open access, the authors have applied a Creative Commons Attribution (CC BY) licence to any Author Accepted Manuscript version arising from this submission. 

\vspace{0.3cm}

{\textbf{Declaration of interests}} \par
The authors report no conflict of interest.
\vspace{0.3cm}

\appendix
\section{Linear stability analysis with free-slip boundary condition}\label{sec:LSA_freeslip}

To investigate the potential impacts of the base flow shape on the instabilities, we perform a LSA with a $\tanh$-shape density (\ref{eq:LSA_R}) and velocity base
\begin{equation}
    U(z)=\gamma\tanh(\iota z),
\end{equation}
where $\iota=1.5\kappa\sqrt{Pr}$ defines the thickness of velocity base. A free-slip boundary condition for velocities is also adopted to understand the effects of boundary conditions.

\begin{figure}
	\centering		
	\includegraphics[width=.7\linewidth, trim=0mm 0mm 0mm 0mm, clip]{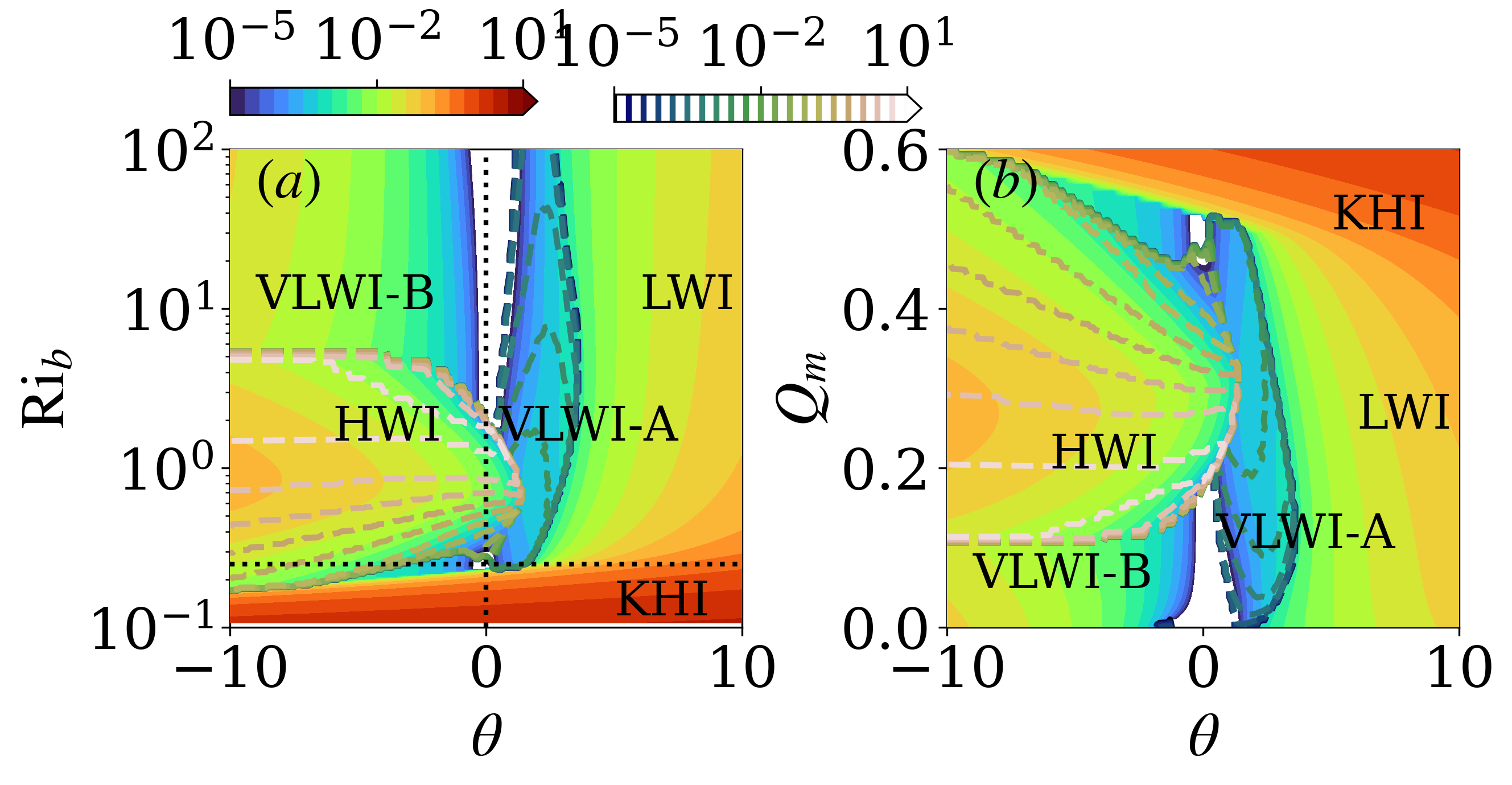}

	\caption{Projection of the fastest growing mode onto parameter spaces: (a) $\mathrm{Ri}_c-\theta$ and (b)$Q_m-\theta$. Solid and dashed lines are the growth rate and frequency, respectively. }
	\label{fig:Qm_ric_theta_freeslip}
\end{figure}

In figure~\ref{fig:Qm_ric_theta_freeslip}, we show the $Ri_b-\theta$ and $Q_m-\theta$ parameter space of the fastest growing modes of the above LSA. Clearly, the five families of instabilities appear even with the different base flow and boundary conditions. It means that these instabilities are not a consequence of an arbitrary flow condition that is  subject to a certain base flow or boundary condition, but rather general flow instabilities that can appear in a wide range of stratified flow systems. Features of these instabilities, e.g. the wave speed, wavelength, growth rate, and regime, are largely consistent with the main cases discussed in \S\ref{sec:param_space}, which suggests, again, the universal features of these instabilities.

\begin{figure}
	\centering		
	\includegraphics[width=.9\linewidth, trim=0mm 0mm 0mm 0mm, clip]{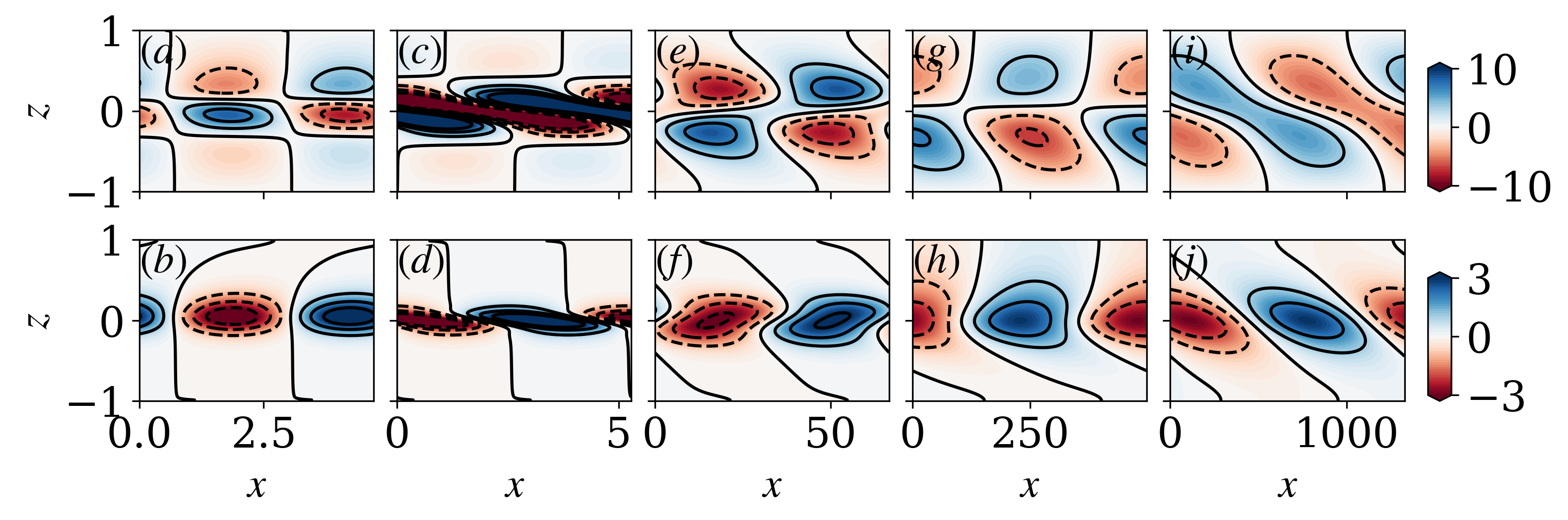}
    \caption{vorticity (1st row) and density (2nd row) eigenfunctions of the fastest growing modes of (a-b) I, HWI, (d-f) II, KHI, (g-i) III, LWI, (j-l) IV, VLWI-DS, and (m-o) V, VLWI-US. }
	\label{fig:eigvec_freeslip}
\end{figure}

\Cref{fig:eigvec_freeslip} show the eigenfunctions of fastest growth modes of the typical case of each instability (marked in figure~\ref{fig:Qm_ric_theta_freeslip}). Note again that the forms of eigenfunctions of each instability are generally consistent with the main cases in \S\ref{sec:eigfun} in the middle region of the channel. However, those intense regions near the wall do not appear with a free-slip velocity boundary condition. Therefore, these near-wall structures as well as the no-slip boundary conditions are not essential to these instabilities.



\bibliographystyle{jfm}
\bibliography{main.bib}

\end{document}